\pdfoutput=1
\documentclass[%
 reprint,
superscriptaddress,
amsmath,amssymb,
aps,
]{revtex4-2}
\pdfoutput=1
\usepackage{graphicx}
\usepackage{dcolumn}
\usepackage{bm}
\usepackage{parskip}

\usepackage{acro}
\usepackage{textcomp, gensymb}
\usepackage{afterpage,natbib,lipsum}
\usepackage{float}
\usepackage{placeins}
\usepackage{xcolor}
\usepackage[none]{hyphenat}
\graphicspath{ {Paper/images/} }

\usepackage{hyperref}

\DeclareAcronym{ssDNA}{
	short = ssDNA ,
	long = single-stranded DNA}
\DeclareAcronym{dsDNA}{
	short = dsDNA ,
	long = double-stranded DNA}
\DeclareAcronym{GUV}{
	short = GUV ,
	long = giant unilamellar vesicle}
\DeclareAcronym{TBE}{
	short = TBE ,
	long = Tris-borate-EDTA}
\DeclareAcronym{TE}{
	short = TE ,
	long = Tris-EDTA}
\DeclareAcronym{DOPC}{
	short = DOPC ,
	long = {1,2-Dioleoyl-\linebreak[0]\textit{sn}-glycero-\linebreak[0]3-phophocholine}}
\DeclareAcronym{NBD-PE}{
    short = NBD-PE ,
	long = {1,2-dioleoyl-sn-glycero-\linebreak[0]3-phosphoethanolamine-N-\linebreak[0](7-nitro-2-1,3-benzoxadiazol-4-yl) (ammonium salt)}}
\DeclareAcronym{ITO}{
	short = ITO ,
	long = indium tin oxide}
\DeclareAcronym{PDMS}{
	short = PDMS ,
	long = polydimethylsiloxane}
\DeclareAcronym{HPLC}{
	short = HPLC ,
	long = high performance liquid chromatography}
\DeclareAcronym{NaCl}{
    short = NaCl ,
	long = sodium chloride}
 
\begin{document}
\preprint{APS/123-QED}
\title{Chemotactic crawling of multivalent vesicles along ligand-density gradients}

\author{Hannah Sleath} 
\affiliation{Department of Chemistry, Imperial College London, Molecular Sciences Research Hub, 82 Wood Lane, London W12 0BZ, United Kingdom}
\affiliation{fabriCELL, Imperial College London, Molecular Sciences Research Hub, 82 Wood Lane, London W12 0BZ, United Kingdom}

\author{Bortolo Mognetti}
\email[]{bortolo.matteo.mognetti@ulb.be}
\affiliation{Interdisciplinary Center for
Nonlinear Phenomena and Complex Systems, Université
Libre de Bruxelles (ULB), B-1050 Brussels, Belgium}

\author{Yuval Elani} 
\email[]{y.elani@imperial.ac.uk}
\affiliation{Department of Chemical Engineering, Imperial College London, Imperial College Road, London SW7 2AZ, United Kingdom}
\affiliation{fabriCELL, Imperial College London, Molecular Sciences Research Hub, 82 Wood Lane, London W12 0BZ, United Kingdom}

\author{Lorenzo Di Michele} 
\email[]{ld389@cam.ac.uk}
\affiliation{Department of Chemical Engineering and Biotechnology, University of Cambridge, Philippa Fawcett Drive, Cambridge CB3 0AS, United Kingdom}
\affiliation{Department of Chemistry, Imperial College London, Molecular Sciences Research Hub, 82 Wood Lane, London W12 0BZ, United Kingdom}
\affiliation{fabriCELL, Imperial College London, Molecular Sciences Research Hub, 82 Wood Lane, London W12 0BZ, United Kingdom}


\begin{abstract}
Living cells are capable of interacting with their environments in a variety of ways, including cell signalling, adhesion, and directed motion. These behaviours are often mediated by receptor molecules embedded in the cell membrane, which bind specific ligands. Adhesion mediated by a large number of weakly binding moieties---multivalent binding---is prevalent in a range of active cellular processes, such as cell crawling and pathogen-host invasion. In these circumstances, motion is often caused by gradients in ligand density, which constitutes a simple example of chemotaxis.  To unravel the biophysics of chemotactic multivalent adhesion, we have designed an experimental system in which artificial cell models based on lipid vesicles adhere to a substrate through multivalent interactions, and perform chemotactic motion towards higher ligand concentrations.  Adhesion occurs \emph{via} vesicle-anchored receptors and substrate-anchored ligands, both consisting of synthetic DNA linkers that allow precise control over binding strength. Experimental data, rationalised through numerical and theoretical models, reveal that motion directionality is correlated to both binding strength and vesicle size. Besides providing insights into the biophysics of chemotactic multivalent adhesion, our results highlight design rules applicable to the development of biomimetic motile systems for synthetic biology and therapeutic applications.
\end{abstract}

\maketitle


\section{Introduction}
\label{chap_Introduction}

Living cells interact with their environments \emph{via} receptors embedded in their plasma membranes. These receptors can bind to specific ligand molecules, resulting in processes vital to biological function such as adhesion to external surfaces~\cite{ahmad2015review,kam2001cell} and the formation of pseudopodia involved in cell motility~\cite{van2004chemotaxis}. Cell adhesion is often mediated by a large number of molecular bonds between cell-membrane receptors and surface ligands. These multivalent interactions can produce complex and useful emergent behaviours, such as binding super-selectivity~\cite{liu2020combinatorial,linne2021direct}, due to the interplay between enthalpic and configurational effects~\cite{varilly2012general}. Numerous experimental and theoretical studies have been conducted to explore various aspects of multivalent interactions between particles, membranes and surfaces, including the strength and rate of adhesion~\cite{mognetti2019programmable, angioletti2017theory, curk2018design, lanfranco2019kinetics, amjad2017membrane}; the growth, size and stability of the contact region between adhering objects~\cite{shenoy2005growth,parolini2015volume}; the self-assembly or fusion of colloidal particles~\cite{beales2007specific,beales2011specific,beales2014application,rogers2016using, nykypanchuk2008dna,flavier2017vesicle,chan2009effects}; and receptor-mediated endocytosis~\cite{di2018steric}.

Cells and viruses adhering through multivalent interactions are known to perform directed motion. Examples include the rolling and subsequent adhesion of leukocytes to inflamed endothelia~\cite{ley2007getting,muller2002leukocyte}; and the diffusion of the Influenza A~\cite{de2020influenza,vahey2019influenza} and Herpes~\cite{delguste2018multivalent} viruses along the cell membrane prior to invasion.

In many instances, motion directionality in multivalent adhesion systems is a result of gradients in the type and density of ligands~\cite{basara1985stimulation,thibault2007fibronectin,haessler2011dendritic}, making it a form of chemotaxis conceptually analogous to the directional swimming of algae, bacteria and sperm cells~\cite{adler1966chemotaxis,choi2016quantitative,eisenbach2006sperm}. Despite relevant studies on particle drift induced by shear flow~\cite{hammer1987dynamical,dasanna2017rolling,porter2021shear,hamming2023receptor} and motion occurring in low valency systems~\cite{perl2011gradient,chang2007directional}, the biophysics of directed crawling mediated by multivalent interactions remains largely unexplored. In particular, the relative importance of factors such as binding strength and the size of the adhering cell or particle on the ability to perform chemotactic crawling is yet to be clarified.

\begin{figure*}[!ht]
    \includegraphics[width=1\textwidth]{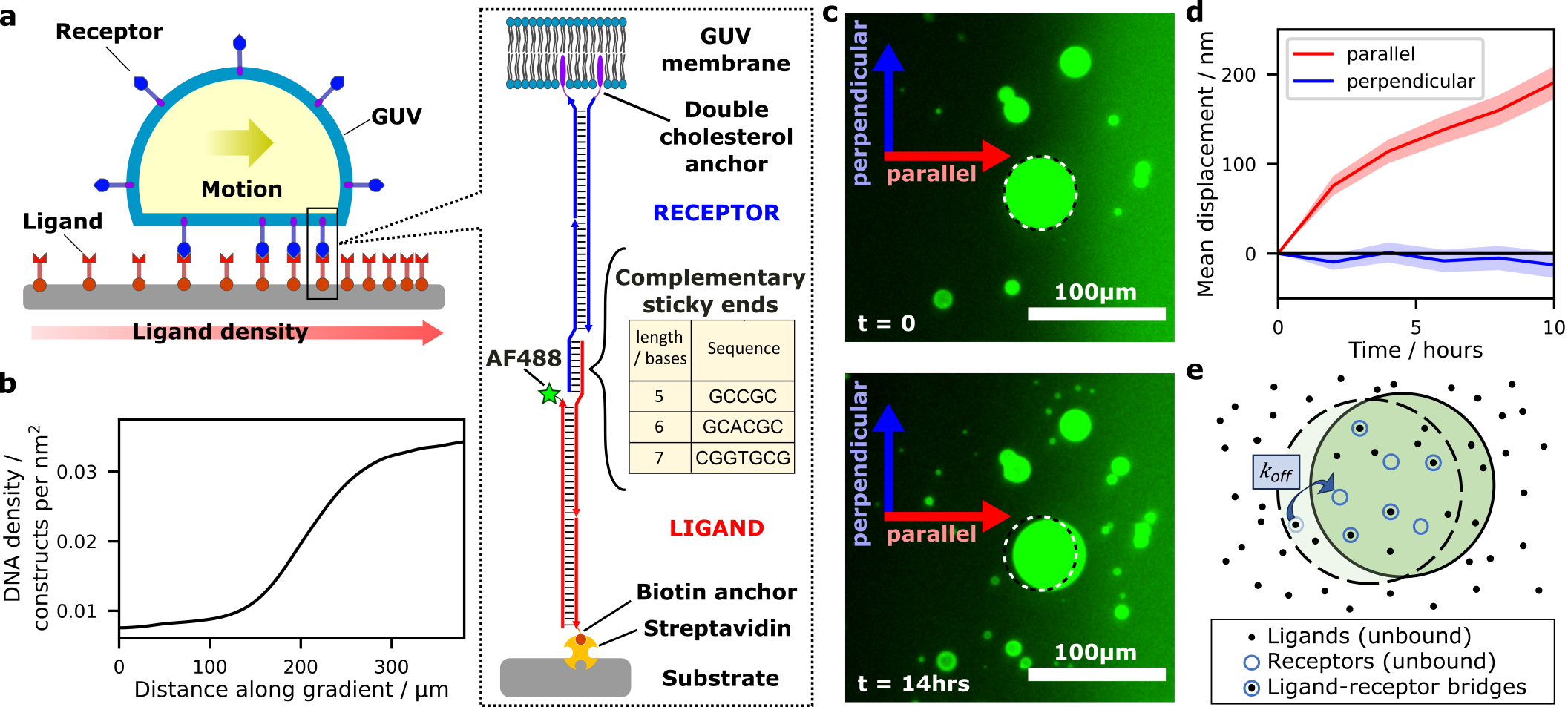}
    \caption{\textbf{Overview of the experimental system.} (a) Schematic of the experimental system, in which a giant unilamellar vesicle (GUV) functionalised with DNA ``receptor" constructs interacts with a surface density gradient of complementary DNA ``ligand" constructs attached to a substrate. Double-cholesterol anchors are used for anchoring the constructs to the GUVs, while biotin-streptavidin connections are used for the substrate. (b) Example of a ligand-density gradient profile from experiments. (c) Microscopy images showing the displacement of a GUV over 14 hours; the original position of the vesicle is marked with a dashed circle. Axes indicating the directions parallel and perpendicular to the gradient are included. (d) Mean displacement over time of vesicles in systems with sticky end length $l=$~5\,nt. Data for the mean displacement parallel and perpendicular to the ligand-density gradient are shown with the standard error of the mean shaded. Positive values of the displacement parallel to the gradient indicate motion towards higher ligand-density regions. (e) Schematic of the theoretical model, illustrating the motion of a GUV over a surface functionalised with ligands. The GUV is attached to the surface by receptor-ligand bridges, which constrain its position. Upon spontaneous unbinding of a bridge near its perimeter, the GUV can explore beyond its original constrained region.}
    \label{fig:Figure1}
\end{figure*}

Here, we address these questions using a combination of theory, numerical modelling, and experiments performed on a biomimetic model system. For experiments, we consider synthetic cellular mimics consisting of \acp{GUV}, interacting with a solid surface through multivalent adhesion. As schematised in Fig.~\ref{fig:Figure1}a, both the ``receptors" on the \acp{GUV} and the ``ligands" on the surface consist of synthetic DNA constructs interacting through selective base-pairing interactions that allow us to precisely modulate ligand-receptor affinity~\cite{mognetti2019programmable,shimobayashi2015direct,parolini2015volume}. A gradient in the surface density of the ligands is established, which causes the artificial cells to drift towards higher ligand concentrations thanks to the reversibility of the ligand-receptor interactions. We find that the vesicle drifting velocity is approximately proportional to the unbinding rate of the ligand-receptor bridges, in agreement with theoretical considerations. We further explore the relationship between \ac{GUV} size and motion, observing a positive correlation between vesicle size and drifting velocity. Coarse-grained simulations based on the model developed in Ref.~\cite{lowensohn2022sliding} produce consistent trends in vesicle motion when varying ligand-receptor affinity and vesicle size.
 
These results provide a deeper insight into the biophysics of multivalent chemotactic motion, which could help to rationalise biological processes with relevance to immunity, host-pathogen interactions and tissue dynamics.\cite{ley2007getting,de2020influenza,basara1985stimulation} Our work also provides design principles for engineering artificial cell chemotaxis in the context of bottom-up synthetic biology, to generate phenomena such as coordinated motion in multicellular systems, or targeting and chasing of specific objects or chemical signals~\cite{blain2014progress,guindani2022synthetic,siton2016toward}. This could lead to development of synthetic cellular solutions valuable in a variety of applications \emph{e.g.} therapeutics and targeted drug delivery~\cite{fischbach2013cell,joseph2017chemotactic,kershaw2002redirecting,moon2011expression,pockaj1994localization,karp2009mesenchymal}.

\section{Results and Discussion}
\label{chap_Results}

\subsection{Experimental system}
A schematic of the experimental system is shown in Fig.~\ref{fig:Figure1}a. Electroformed \acp{GUV} prepared from 1,2-Dioleoyl-sn-glycero-3-phosphocholine (DOPC), with typical diameters of 1--50\,$\mu$m (see Fig~S1), are functionalised with DNA constructs, here referred to as ``receptors''. The receptors feature a double cholesterol anchor that irreversibly partitions within the lipid bilayer.~\cite{mognetti2019programmable,pfeiffer2004bivalent} As DOPC bilayers are fluid, receptors can freely diffuse laterally and redistribute across the surface of the GUV. The surface density of receptors is approximately 0.008\,nm$^{-2}$ (see Materials and Methods, Section~\ref{chap_Methods}). The receptors interact with a second set of DNA constructs, here referred to as ``ligands'', which feature a biotin moiety for anchoring to a streptavidin-coated substrate. The receptor and ligand constructs feature complementary \ac{ssDNA} sticky ends with lengths $l$ ranging from five to ten nucleotides (nt), which can reversibly bind to each other. The constructs also feature rigid \ac{dsDNA} spacers  designed to control their spatial extent, and short poly-T domains to increase the flexibility and configurational freedom~\cite{mognetti2019programmable}. The DNA sequences are listed in Tab.~S1 with schematics of the constructs in Fig.~S2. We verified the correct assembly of the constructs and the binding of complementary receptor-ligand pairs using agarose gel electrophoresis (see Figs~S3 and S4), while fluorescence microscopy was used to confirm attachment of the receptor constructs to the vesicle membranes (see Fig.~S5).

We set up a surface density gradient of ligands on the substrate, with density typically varying from  0.01\,nm$^{-2}$ to 0.03\,nm$^{-2}$ over a distance of 100\,$\mu$m (see Fig.~\ref{fig:Figure1}b). The method for generating the gradient is outlined in the Materials and Methods (Section~\ref{chap_Methods}), and images of typical ligand-density gradients are shown in Fig.~S6. Receptor-functionalised \acp{GUV} are then deposited onto the gradient region where they adhere to the surface as sketched in Fig.~\ref{fig:Figure1}a. For sufficiently strong adhesion, the \acp{GUV} take the shape of a truncated sphere, forming a flat adhesion patch visible in confocal cross sections (see Fig.~S7)~\cite{mognetti2019programmable,shimobayashi2015direct,amjad2017membrane}. Time lapse epifluorescence microscopy videos are recorded, where both the \acp{GUV} and the ligand-gradient can be visualised thanks to calcein dye loaded in the vesicles and Alexa Fluor 488 modifications on a sub-set of ligands. Example images in Fig.~\ref{fig:Figure1}c show a \ac{GUV} migrating in the direction of the ligand-density gradient. Images are computationally segmented to identify vesicle trajectories (see SI Section~I). Projecting \ac{GUV} displacement onto the directions parallel and perpendicular to the local gradient direction (see SI Section~I\,C) allows us to quantify motion directionality, as exemplified in Fig.~\ref{fig:Figure1}d.

\subsection{Numerical modelling}
\label{subsec:theory_and_simulations}
Coarse-grained simulations based on Ref.~\cite{lowensohn2022sliding} are used to computationally characterise the multivalent chemotactic system. We map the vesicles onto 2D rigid disks of radius $R$, representing the perimeter of the flat adhesion patch, as shown in Fig.~\ref{fig:Figure1}e. In the following, we refer to the simulated objects as vesicles or disks, interchangeably. Thermal fluctuations in vesicle shape are neglected, and the disk-surface distance maintained constant. To keep simulations affordable, we use disks with diameter $2R=0.2\,\mu$m, 1$\,\mu$m, and $1.8\,\mu$m, smaller by a factor $\sim 10$ compared to the \acp{GUV} used in experiments. The surface is randomly decorated by ligands to generate a linear density profile with slope $\lambda$. The average surface density of the ligands at the starting location of the disks is set to $\rho_\mathrm{L}=0.021\,$nm$^{-2}$, while the average density of receptors on the vesicles is $\rho_\mathrm{R}=0.008\,$nm$^{-2}$, consistently with the nominal experimental values. The simulated gradient slope is set to $\lambda=\rho_\mathrm{L}/10\, \mu$m$^{-1}$, an order of magnitude higher than experimental gradients such that the steepness of the gradient relative to vesicle size is maintained. Note that further decreasing $\lambda$ would rapidly make the system computationally untreatable, as for milder gradients the drifting component of the motion is overwhelmed by the stochastic one. Receptors on the vesicle can form bridges by binding free ligands within the projection of the disk onto the surface. The rate constants at which bridges form and break are given, respectively, by~\cite{mognetti2019programmable}

\begin{eqnarray}
\label{eq:kon_koff}
k_\mathrm{on} = \frac{\alpha\,k_\mathrm{on}^\mathrm{sol}}{3 \pi R^2 L}
& \qquad &
k_\mathrm{off} = \alpha\,\rho_0\,k_\mathrm{on}^\mathrm{sol}\exp[\beta \Delta G_0],
\end{eqnarray}

where $k_\mathrm{on}^\mathrm{sol}$ is the hybridisation rate of free oligonucleotides in solution, $L$ is the length of the \ac{dsDNA} spacer of the ligands and receptors, $\Delta G_0$ is the standard hybridisation free energy of the sticky ends, and $\rho_0=1$\,M is the standard concentration. $\alpha$ is a non-dimensional factor $<1$, accounting for the fact that the hybridization kinetics are expected to be slower as a result of the DNA constructs being tethered to surfaces. We set $\alpha=0.1$ as an estimate based loosely on observations made in Ref.~\cite{linne2021direct}. As simulated and theoretical drift velocities are proportional to $\alpha$, changing this parameter affects the absolute values but not the trends generated by varying other system parameters. Vesicles are represented as hemispheres with a total surface area equal to $3\pi R^2$. Accordingly, each free sticky end is taken as uniformly distributed within a layer of thickness equal to the ligand/receptor length ($L$) surrounding the vesicle's surface, and a resulting density equal to $1/(3\pi R^2 L)$. This density is used to calculate the binding rate $k_\mathrm{on}$ using standard reaction equations \cite{mognetti2019programmable}. By setting $k_\mathrm{on}^\mathrm{sol}=10^6$\,M$^{-1}$s$^{-1}$ (an approximate estimate for short DNA oligomers~\cite{zhang2009control}), $L=10$\,nm, and $\Delta G_0$ values estimated using the nearest neighbour thermodynamic model~\cite{markham2005dinamelt,markham2008unafold,dimitrov2004prediction,santalucia1998unified}, we obtain $k_\mathrm{off}=$~1.10\,s$^{-1}$, 0.174\,s$^{-1}$, and 0.0084\,s$^{-1}$ for sticky ends with $l=
$~5, 6, and 7\,nt, respectively.

In the model, we do not track the specific position of the receptors when free (\emph{i.e.} not bound to a ligand), consistently with the observation that the receptors rearrange onto the bilayer much faster compared to the timescales of vesicle motion. At any point in time, the set of configurations available to the vesicle is limited by the formed bridges,  as they are constrained to remaining inside the perimeter of the disk. Until the bonds rearrange, the disk can thus only rattle around the small sub-set of $n_\mathrm{cb}$  ``constraining bridges'' that can make direct contact with its perimeter (where $\langle n_\mathrm{cb} \rangle= 5$~\cite{lowensohn2022sliding}). Motion of the vesicle over larger distances can only emerge following reconfiguration of the ligand-receptor bridges, which thus limits motility~\cite{lowensohn2022sliding}.

We simulate the system using a reaction-diffusion algorithm~\cite{mognetti2019programmable,jana2019translational}. At each simulation step, of duration $\Delta t$, we iteratively employ the Gillespie algorithm~\cite{gillespie1977exact} to break existing ligand-receptor bridges or form new ones, and compute the time required for each reaction, until the total reaction time exceeds $\Delta t$. Following the reaction step, we update the position of the disk by randomly (uniformly) selecting a new centre of mass from all the possible locations compatible with the new configuration of bridges~\cite{lowensohn2022sliding}. The simulation code can be found online~\cite{simulationcodegithub}.

\subsection{Weaker binding enhances directional motion}
\label{subsec:stickyendlength}
\begin{figure*}[!ht]
    \includegraphics[width=1\textwidth]{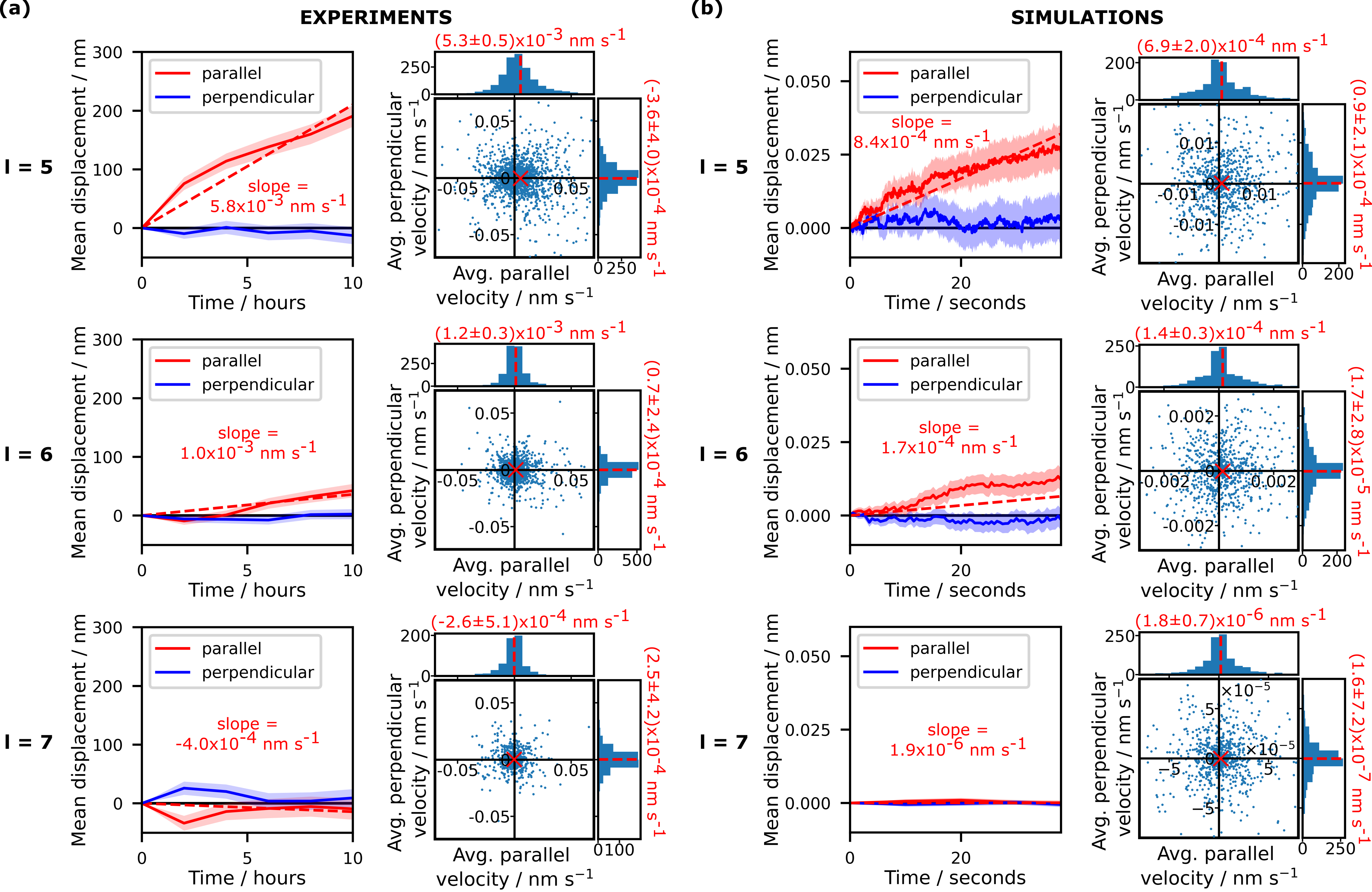}
    \caption{\textbf{Drift velocity increases with decreasing sticky end length.} Data illustrating the effect of sticky end length $l$ on the motion of the GUVs along the ligand surface density gradient, for systems with $l=$~5--7\,nt. Experimental data for vesicles of all sizes are shown in (a), while simulation data for 1\,$\mu$m-diameter vesicles are shown in (b). Solid lines indicate the mean displacement of the vesicles parallel and perpendicular to the gradient over time, where positive values of velocity parallel to the gradient indicate motion towards regions of higher ligand density. Dashed lines indicated linear fits to the data for parallel motion, with the slope annotated.  Simulated trajectories have been cropped to the same duration to enable visual comparison, while the straight lines have been fitted to the entire, uncropped trajectories (displayed in Figs~S9 and S13). The scatter plots with marginal histograms show the average velocities of individual vesicles parallel and perpendicular to the gradient, calculated as total trajectory displacement divided by duration. The y-axes of the histograms indicate the number of vesicles in each bin. Red dashed lines and the red cross mark the mean of the distributions. Note that the axes have been cropped to aid visualisation of the majority of data points. Graphs including all data points are shown in Figs~S10 and S14.}
    \label{fig:Figure2_exp_and_sim}
\end{figure*}

A key factor affecting the strength of multivalent binding and the corresponding vesicle mobility is the strength and (un)binding rates of individual receptor-ligand interactions. This factor can be tuned by varying the length of the complementary sticky ends; in our experiments and simulations we studied systems with sticky end lengths $l=$~5, 6 and 7\,nt.

Experimentally, under all these conditions, the vesicles adhere to the substrate and form a stable, flat adhesion patch, as confirmed by confocal cross sections (see Fig.~S7). As discussed in Section~I\,D of the SI, in all samples, we notice that a small proportion of vesicles (approximately 5--6\%) remains much more static that the rest of the population, indicated by a significantly smaller diffusion coefficient (see Fig.~S8). We believe that this effectively immobile sub-population could emerge due to a number of factors, such as sample impurities or surface defects being erroneously identified as vesicles; non-specific adhesion or trapping of the vesicles on the substrate; or tracking inaccuracies. In the following data analysis the immobile population is excluded, although all plots in the main text have been replicated in the SI with the immobile population included for transparency (Figs~S9--14), showing no qualitative changes to the findings outlined below.

To study the speed and direction of the vesicle motion in both experiments and simulations, each trajectory was separated into components parallel and perpendicular to the local density gradient of DNA ligand constructs. Experimental and simulated data are collated in Fig.~\ref{fig:Figure2_exp_and_sim}a and \ref{fig:Figure2_exp_and_sim}b. Here we report the mean displacement projected along the directions parallel and perpendicular to the ligand-density gradient (left) and the two-dimensional distributions of average velocities, as computed from the initial and final positions of the vesicles in the trajectories (right). From the average velocity distributions, we note that, in all cases, the motion is predominantly stochastic. However, while the projection of the average velocity onto the direction perpendicular to the gradient is centered around zero, \emph{i.e.}, shows no directional bias, the projection along the direction of the gradient has non-zero mean for $l=$~5 and 6\,nt, both in experiments and simulations. The observed bias indicates motion towards denser regions of the ligand carpet, as expected. For $l=7$\,nt, no clear directional bias is noted, in simulations or experiments. The width of the average velocity distributions also decreases with increasing $l$, indicating that vesicles with shorter sticky ends diffuse less, regardless of directionality, as shown in Fig.~S15. This relative decrease is more pronounced in simulations than it is in experiments, as expected given that experimental data are inevitably impacted by static localisation errors.

Directional bias of the motion towards higher ligand densities is better visualised from the mean displacement data shown as a function of time (Figs~\ref{fig:Figure2_exp_and_sim}a and \ref{fig:Figure2_exp_and_sim}b, left). In all cases, data on the motion perpendicular to the gradient average to zero (within standard error), while data on the parallel motion display a non-zero slope for $l=5$ and 6\,nt, with the latter being less pronounced. For $l=7$\,nt, mean displacement parallel to the gradient shows a negligible slope.

From the scatter plots in Fig.~\ref{fig:Figure2_exp_and_sim} we calculate the average experimental drift velocities for the $l=$~5 and 6\,nt systems to be $5.3\times 10^{-3}$\,nm\,s$^{-1}$ and $1.2\times 10^{-3}$\,nm\,s$^{-1}$ respectively. Over experimental timescales (\emph{i.e.} 10 hours), this equates to total displacements of 190\,nm and 43\,nm respectively, which are several orders of magnitude smaller than the typical vesicle diameter ($\approx 10$\,$\mu$m; see Fig.~S1). However, we note that the distribution of vesicle displacements parallel to the gradient has a positive skew for both $l=$~5 and 6\,nt, with maximum displacements along the gradient of 5.8\,$\mu$m and 5.1\,$\mu$m respectively, comparable with the average vesicle radius. For comparison, a freely diffusing vesicle of this size would travel $\approx$~80\,$\mu$m along the surface in this time scale (using the Stokes-Einstein equation to estimate the diffusion constant), which shows that the vesicle mobility is reduced by the multivalent adhesion.

To put the extent of the vesicle displacements into perspective, we consider the spacing between ligand-receptor bridges. In Fig.~S16 we see that the system is in a saturation regime where the ligand-receptor bridge density is limited by either the density of ligands or the density of receptors. This means that the bridge density is on the order of 0.01\,nm$^{-2}$, which corresponds to an inter-bridge spacing of 10\,nm. A vesicle displacement of $\sim5\,\mu$m is several orders of magnitude higher than the inter-bridge spacing, indicating that the observed motion cannot simply be due to the rattling of vesicles around fixed bridging points but must require the formation and breakage of a large number of bridges.

Furthermore, we note that a typical vesicle with a diameter of 10\,$\mu$m, located centrally in the gradient region of Fig.~\ref{fig:Figure1}b, spans ligand densities that only differ by $\sim10$\% (from 0.019\,nm$^{-2}$ to 0.021\,nm$^{-2}$). Therefore, during each formation of a new ligand-receptor bridge, it is only 10\% more likely (at most) for the bridge to form on the side of the contact region facing towards higher ligand density, compared to the opposite side. This small difference in binding probability across the span of the vesicle suggests that a large number of bridge formation and breakage events are required for the directional vesicle drift along the gradient to be distinguishable from  stochastic motion.

When comparing experiments with simulated trajectories, we note that the average drift velocities from experimental data for $l=$~5 and 6\,nt are an order of magnitude higher than the simulated velocities (see Fig.~\ref{fig:Figure2_exp_and_sim}). This deviation could be due to the simulated vesicle diameters being an order of magnitude smaller than in experiments, although the simulated gradients have been adjusted to achieve the same gradient steepness relative to vesicle size. The effect of vesicle size on drift velocity is further discussed in Section~\ref{subsec:vesiclesize}. The difference in drift velocities between experiments and simulations could also be due to differences in the gradient profiles or to small convective flows  and tracking inaccuracies inflating experimental estimates. Furthermore, calculations of the bridge formation and breakage rates (Eq.~\ref{eq:kon_koff}) for the simulations involve assumption of the correction factor $\alpha$, which could introduce further discrepancies.

Besides systems with sticky end lengths $l=$~5--7\,nt, we also experimentally tested samples with shorter sticky end lengths ($l=$~3 and 4\,nt), with the goal of further increasing vesicle mobility. However, we found that the binding strength was insufficient to keep the \acp{GUV} adhered to the surface, causing them  to drift driven by  convection or gravity (see SI Fig.~S17). We additionally designed constructs with longer sticky ends ($l=$~8--10\,nt), which showed negligible drift consistent with the data in  Fig.~\ref{fig:Figure2_exp_and_sim} for $l=7$\,nt.

\subsubsection{Comparison of experimental drift velocities with theory}

For a given number of ligands and receptors, the drift velocity is expected to be a function of the off rate ($k_\mathrm{off}$), the number of bridges ($n_\mathrm{b}$), and the diffusion constant of free (non-adhering) vesicles ($D$). In systems featuring many bridges or low unbinding rates, the displacement dynamics of the vesicles is limited by the timescales of ligand/receptor reactions. This can be verified by comparing two characteristic timescales. The first  is the typical timescale through which the translational configurational space available to the vesicles changes following a binding and unbinding of ligands/receptors, which can be estimated as $\tau_\mathrm{r}=1/(2 k_\mathrm{off} n_\mathrm{cb})$, with $\langle n_\mathrm{cb}\rangle \approx5$~\cite{lowensohn2022sliding}. The second is the diffusion timescale needed by a vesicle to explore the available space: $\tau_\mathrm{d}=\pi R^2 n_\mathrm{cb} / (D n_\mathrm{b}^2)$.

In conditions considered here, $\tau_\mathrm{r}$ is at least 7 orders of magnitude bigger than $\tau_\mathrm{d}$ (see SI Section~II\,A), making the dynamics of our systems reaction-limited.  Under these circumstances, we expect $v_\mathrm{drift}$ to be entirely determined by $k_\mathrm{off}$, resulting in the following scaling  
\begin{equation}
    v_\mathrm{drift} \propto k_\mathrm{off} f(R,n_\mathrm{b}),
\label{Eq:vdrift_prop}
\end{equation}
where $f(R,n_\mathrm{b})$ is a general function on the variables $R$ and $n_\mathrm{b}$.

For comparison, we derive an analytical prediction of the drift velocity. In our previous contribution, we predicted that a receptor-decorated disk on a uniformly ligand-decorated surface would diffuse with $D \sim k_\mathrm{off} R^2/\langle n_\mathrm{b}\rangle^2$~\cite{lowensohn2022sliding}. By applying the fluctuation-dissipation theorem to this expression, we find  (see SI Section~II\,B)
\begin{equation}
    v_\mathrm{drift}^\mathrm{FD} = -\beta D \frac{\mathrm{d} V_\mathrm{multi}}{\mathrm{d}x} = -\frac{ \lambda k_\mathrm{ off} }{2\langle\rho_\mathrm{b}\rangle^2 } \log \left[ 1- \frac{\langle \rho_\mathrm{b}\rangle}{\langle \rho\rangle }\right], 
    \label{Eq:vdrift_FD}
\end{equation}
where $V_\mathrm{multi}$ is the multivalent free-energy~\cite{martinez2014designing,mognetti2019programmable} and $\langle \rho \rangle$ ($\langle \rho_\mathrm{b} \rangle$) the density of (bound) ligands found at the center of the disk. The expression in Eq.~\ref{Eq:vdrift_FD} agrees with the relationship anticipated in Eq.~\ref{Eq:vdrift_prop}. However, the theoretical prediction in Eq.~\ref{Eq:vdrift_FD}, substantially overestimate drift velocities determined in both experiments and simulations, as shown in Figs~S18 and S19. 
The discrepancies between Eq.~\ref{Eq:vdrift_FD} and simulations decrease at low values of $n_\mathrm{b}$. This discrepancy can be rationalised by noting that, for large $n_\mathrm{b}$ (and thus large drift forces), our systems may deviate from the linear response regime under which the fluctuation-dissipation theorem applies. This observation is in agreement with previous findings that coarse-grained dynamics in non-equilibrium conditions are not necessarily described by a simple Langevin equation~\cite{widder2022generalized,netz2023derivation}.

Although the theoretical prediction in Eq.~\ref{Eq:vdrift_FD} does not appear to be valid under experimentally relevant conditions, we still expect the drift velocity to follow the general scaling of Eq.~\ref{Eq:vdrift_prop}. As previously mentioned, Fig.~S16 indicates that for $l=$~5,6 and 7\,nt, the system is in a saturation regime where the bridge density $\langle n_\mathrm{b}\rangle$ is limited by either the ligand density or the receptor density. It follows that the $R$ and $n_\mathrm{b}$ dependencies in Eq.~\ref{Eq:vdrift_prop} should be unchanged when comparing experiments with different $l$, resulting in $v_\mathrm{drift} \propto k_\mathrm{off}$.

To verify the predicted proportionality, we compute the ratio between $v_\mathrm{drift}$ values for systems with different $l$, and compare it with the ratio between $k_\mathrm{off}$ values. From the linear fits in Fig.~\ref{fig:Figure2_exp_and_sim}, we extract $v^{l=5}_\mathrm{drift}/v^{l=6}_\mathrm{drift}=5.84$ and 4.90 for experiments and simulations, respectively. Both these values are in good agreement with $k^{l=5}_\mathrm{off}/k^{l=6}_\mathrm{off}=6.30$, obtained using the values reported in Section~\ref{subsec:theory_and_simulations}. The small discrepancy possibly derives from using the same $\alpha$ for both sticky ends when computing $k_\mathrm{off}$ (Eq.~\ref{eq:kon_koff}), with $\alpha$ being, in principle, sequence dependent. While similar comparisons could not be carried out for the low-mobility system with $l=7$\,nt, the analysis for $l=5$\,nt and $l=6$\,nt supports a scenario where the drift velocity is dictated by unbinding rates, in alignment with the model leading to Eq.~\ref{Eq:vdrift_prop}.

\FloatBarrier
\subsection{Larger \acp{GUV} exhibit chemotaxis to a greater extent}
\label{subsec:vesiclesize}

\begin{figure}[!ht]
    \includegraphics[width=1\columnwidth]{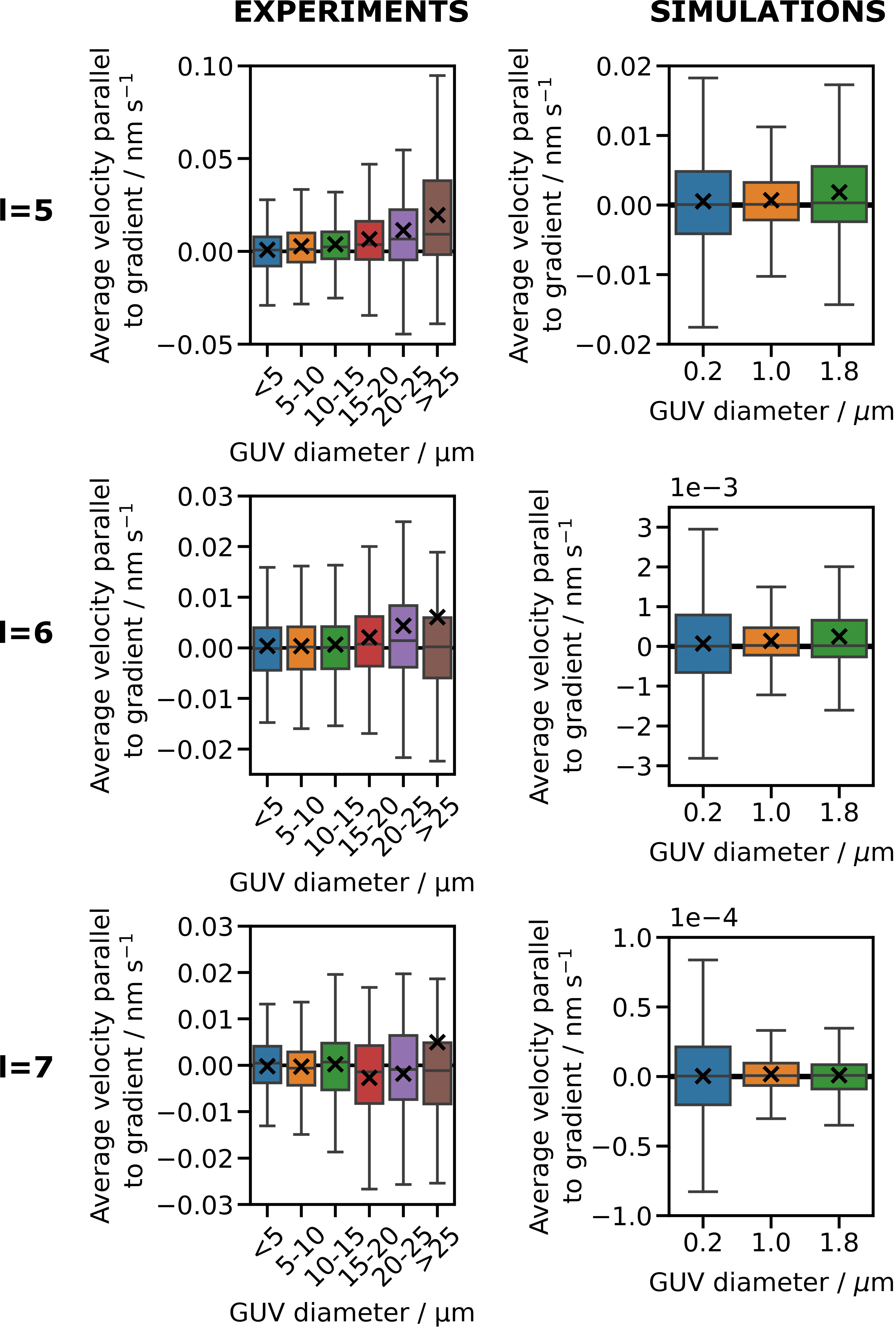}
    \caption{\textbf{Drift velocity increases with increasing vesicle diameter.} Box plots illustrating the relationship between \ac{GUV} size and the average drift velocity of along the ligand density gradient, for sticky length $l=$~5, 6 and 7\,nt. Data show the distribution of average velocity along the gradient, which has been calculated as total displacement parallel to the gradient divided by trajectory duration. Positive values indicate motion of the  towards regions of higher ligand surface density. For each box plot, the mean is marked with a cross. Note that outliers have been excluded for better visualisation. Graphs with outliers included are presented in Fig.~S10 and S14.}
    \label{fig:Figure3}
\end{figure}

We also experimentally observe that the size of the \acp{GUV} has an effect on the drift velocity along the gradient. When binning the experimental trajectories by vesicle diameter, we notice that larger vesicles travel on average a greater distance along the gradient, as shown in Fig.~\ref{fig:Figure3} (left). The trend is most evident in the $l=5$\,nt system, where vesicles are the most mobile, but also present for the less mobile $l=6$\,nt system. Simulated trajectories for vesicles of varying diameters confirm the experimental trends for $l=5$ and 6\,nt (see Fig.~\ref{fig:Figure3}, right). In both the experiments and simulations, we found that the average vesicle drift velocity is typically greater than the median, due to the velocity distribution having a positive skew. We do not observe a clear relationship between vesicle size and motion along the gradient for $l=7$\,nt, in neither experiments nor simulations, likely due to any signal being drowned by noise in this low-mobility system. Regardless of size, we observe that, on average, vesicles drift a negligible distance along the direction perpendicular to the gradient, as shown in Fig.~S12 and S21 and consistent with the data in Fig.~\ref{fig:Figure2_exp_and_sim}.

One possible explanation for the relationship between vesicle size and drift velocity in the mobile systems ($l=$~5 and 6\,nt) is that larger vesicles cover a greater extent of the ligand density gradient, corresponding to a greater change in binding affinity across the width of the vesicle. Following from the discussion in Section~\ref{subsec:stickyendlength}, for a vesicle located centrally on the gradient depicted in Fig.~\ref{fig:Figure1}b, the ligand density varies by $\sim$5\% across a 5\,$\mu$m-diameter contact region, and by $\sim$20\% across a 20\,$\mu$m-diameter contact region. This means that the larger vesicle would have a higher rate of bridge formation on the side of the contact region facing up the gradient compared with the smaller vesicle, resulting in a greater drift velocity. This effect has also been discussed in Ref.~\cite{hamming2023receptor}, where the motility of influenza virus particles adhered to surface-bound molecular density gradients was studied. In this case, no directional bias was observed in the virus motion, which the authors ascribed to the small size of the virus particles relative to the steepness of the ligand-density gradient.

Finally, we note that while larger vesicles can more effectively sense gradients, they are also less mobile due to the larger number of bridges~\cite{lowensohn2022sliding}. These two competing factors perfectly compensate each other in the theoretical prediction reported in Eq.~\ref{Eq:vdrift_FD}, where $v^\mathrm{FD}_\mathrm{drift}$ is not dependent on $R$. Evidence of a size-dependent drift velocity in both experiments and simulations further highlights the limitations of the theoretical approach leading to Eq.~\ref{Eq:vdrift_FD} in the strong-binding regime relevant to our system.

\section{Conclusions}
\label{chap_Conclusions}

In summary, through experiments, theory and simulations, we studied the directional motion of receptor-decorated lipid vesicles adhering to a ligand-decorated surface, in the presence of a gradient in ligand density -- a model system for chemotactic crawling. Experimentally, both receptors and ligands consisted of synthetic DNA constructs, the former connected to giant unilamellar lipid vesicles and the latter to a solid surface. Coarse-grained simulations relied on a multi-scale approach first reported in Ref.~\cite{lowensohn2022sliding}, which accurately describes the binding and unbinding dynamics of ligands and receptors through a Gillespie algorithm.

Both simulations and experiments showed directional motion of the vesicles towards ligand-dense regions of the surface. The magnitude of the directional drift was observed to decrease with increasing ligand-receptor affinity, which could be easily controlled by changing the length of the single-stranded DNA sticky ends through which our constructs interact.

With theoretical arguments we demonstrated that, in the regime relevant to our system, directional motion is limited by the timescales of the ligand-receptor reactions. The corresponding relationship between drift velocity and ligand-receptor unbinding rates were confirmed by both experiments and simulations.

We also observed a positive correlation between vesicle size and drift velocity in both experiments and simulations. This trend, more noticeable in more mobile systems, is explained by the ability of larger vesicles to probe a greater extent of the ligand density gradient, thus generating greater spatial asymmetry in the ligand-receptor binding probabilities.

Our findings offer quantitative insights on the mechanisms underpinning directional motion in multivalent systems, highly relevant to a variety of biological processes involving the membrane interactions. Examples include immune cells adhesion~\cite{ley2007getting,muller2002leukocyte}, viral invasion~\cite{de2020influenza,vahey2019influenza,delguste2018multivalent}, and tissue dynamics~\cite{ley2007getting,de2020influenza,basara1985stimulation}. In future studies, it would be interesting to explore parameter spaces such as temperature, buffer conditions and gradient steepness, as well as working in more weakly-binding regimes (\emph{i.e.} by reducing sticky end length or decreasing DNA density); this may require alteration of the experimental setup to reduce convection effects.

Our models and experimental implementation will also support efforts to engineer directed motion in synthetic cellular systems. Mobile synthetic cells capable of performing simple chemotaxis would indeed be valuable for a vast range of applications, such as: smart drug delivery systems that can target and release drugs at specific locations in the body~\cite{fischbach2013cell,joseph2017chemotactic,kershaw2002redirecting,moon2011expression,pockaj1994localization,karp2009mesenchymal}; vesicle-based biosensors or biomedical imaging systems that signal the presence of specific molecules or conditions~\cite{sforzi2020liposome,liu2013liposomes,mazur2017liposomes}; or tissue engineering applications where gradients of molecules or growth factors are used to direct cell migration and tissue formation~\cite{monteiro2014liposomes,shafiei2021comprehensive,sarkar2019liposome}.

\section{Materials and Methods}
\label{chap_Methods}

\subsection{Experimental materials}

\paragraph{DNA oligonucleotides.} Sequences of the DNA oligonucleotides were designed using the NUPACK design tool~\cite{zadeh2011nupack}. DNA strands with 5'-cholesteryl modifications were purchased from Eurogentec (Liege, Belgium) with \ac{HPLC} purification. All other DNA strands were purchased from Integrated DNA Technologies (Coralville, Iowa, United States). Modified strands were purified via \ac{HPLC}, and unmodified strands were purified by standard de-salting.

\paragraph{Buffers.} NaCl (BioUltra, $>$99.5$\%$) and 100$\times$ \ac{TE} buffer were purchased from Sigma-Aldrich (Gillingham, UK). 10$\times$ \ac{TBE} was purchased from ThermoFisher Scientific (UK). All buffer solutions were diluted with the appropriate amount of MilliQ water, and were filtered through 0.22\,$\mu$m pore poly(ether sulfone) filters (Millex) prior to use.

\paragraph{Gel electrophoresis. }
Agarose was purchased from Sigma-Aldrich (Gillingham, UK). Ultra Low Range DNA Ladder and 6$\times$ TrackIt Cyan/Yellow Loading Buffer were purchased from ThermoFisher Scientific (UK). 10,000$\times$ SYBR Safe DNA gel stain was purchased from APExBIO Technology LLC (Houston, Texas, USA).

\paragraph{Vesicle generation.} D-(+)-glucose ($>$99.5$\%$), sucrose, chloroform and indium tin oxide coated glass slides (surface resistivity 15-25\,$\Omega$\,sq$^{-1}$) were purchased from Sigma-Aldrich (Gillingham, UK). $>$99$\%$ \ac{DOPC} and \Ac{NBD-PE} was purchased from Avanti Polar Lipids Inc. (Alabaster, Alabama, USA).

\paragraph{Well plates and adhesive covers.} Streptavidin-coated high capacity 96-well strip plates were purchased from Sigma-Aldrich (Gillingham, UK). Adhesive clear foils for 96-well plates were purchased from Sarstedt Ltd (Leicestershire, UK).

\subsection{Experimental methods}
\label{subsec_exptmethods}
\paragraph{DNA Reconstitution.}
The DNA strands were shipped lyophilised and were reconstituted to a concentration of approximately 100\,$\mu$M in 1$\times$ \ac{TE} buffer (1\,mM EDTA, 10\,mM Tris, pH\,8.0). The concentration of the reconstituted DNA was determined via UV-vis spectrophotometry on a Thermo Scientific NanoDrop One, by measuring the ratio of absorbance at 260\,nm to the sequence-specific extinction coefficient of the DNA. Stock solutions of DNA were then stored at $-$20$\degree$C.

\paragraph{DNA Hybridisation.}
Assembly of the multistranded receptor and ligand constructs was facilitated by thermal annealing. For each construct, the constituent strands were diluted to 2\,$\mu$M and combined in TE buffer containing 100\,mM NaCl. The strands were first heated to 90$\degree$C for 5 minutes to ensure melting of all DNA duplexes, and then cooled to 20$\degree$C at a rate of $-$0.5$\degree$C min$^{-1}$, on a Bio Rad C1000 Thermal Cycler. The sizes of the annealed DNA constructs were evaluated \emph{via} agarose gel electrophoresis to verify correct folding, as reported in Figs~S3 and S4.

\paragraph{Agarose gel electrophoresis.}
3\% (w/v) agarose gels were casted in 1$\times$ \ac{TBE} buffer with 1$\times$ SYBR Safe DNA gel stain. The 10\,$\mu$l wells were loaded with approximately 300\,ng DNA samples with loading dye; the outer two wells were loaded with DNA ladders, to enable comparison with the DNA samples. The gels were run for 90 minutes at 120\,V (electric field strength of 6\,V\,cm$^{-1}$) and then imaged using a Syngene Dyversity 4 gel imager.

\paragraph{Microscopy.}
Epifluorescence microscopy imaging was performed on a Nikon Eclipse Ti2-E inverted microscope using a Nikon CFI Plan Apochromat Lambda D 10$\times$ dry objective (NA 0.45). Confocal microscopy imaging was performed on a Leica TCS SP5 Confocal microscope using a HCX PL APO CS 63.0$\times$ oil-immersion objective (NA 1.40).

\paragraph{Preparing vesicles.}
The \acp{GUV} were produced \emph{via} electroformation, a straightforward and commonly used method for vesicle generation~\cite{angelova1986liposome}.
Lipid solutions consisting of \ac{DOPC} with 1\,mol$\%$ fluorescent lipid \ac{NBD-PE} were prepared by dissolving the appropriate quantities of lipids in chloroform to yield a 1 mg ml$^{-1}$ solution. 30\,$\mu$l of this solution was spread evenly on an \ac{ITO} slide and vacuum-desiccated for 30\,minutes to remove residual chloroform, resulting in a lipid film. A 5\,mm-thick \ac{PDMS} spacer with a central cut-out was sandwiched between this film and another \ac{ITO} slide to create a chamber, and the chamber was filled with a solution consisting of 300\,mM sucrose and 50\,$\mu$M calcein in water. A function generator (Aim-TTi, TG315, Huntingdon, UK) was used to apply an alternating electric field at a peak-to-peak voltage of 1.5\,V was applied across the \ac{ITO} slides at 10\,Hz for two hours, followed by one further hour at 2\,Hz. During this time, the chamber was left in a 60$\degree$C oven to ensure that the lipid was in the fluid phase. Finally, the chamber was opened and the resulting vesicles were collected. The vesicle sample was inspected \emph{via} epifluorescence microscopy, by exciting both the \ac{NBD-PE} fluorophore and the encapsulated calcein. Note that the calcein signal overwhelms the signal from the \ac{NBD-PE}; the latter was included to aid visualisation of the vesicles in the case of any calcein leakage. Ease of imaging can be improved by mixing the vesicle sample with a solution of 300\,mM glucose in MilliQ water, such as in a 1:10 ratio of vesicles to glucose solution. This introduces a density difference which causes the vesicles to settle at the bottom of the viewing chamber, and a refractive index contrast to enable viewing in brightfield. The size distributions of vesicles in our experiments are shown in Fig.~S5.

\paragraph{Vesicle functionalisation.}
\Acp{GUV} were functionalised by incubation with cholesterolised DNA constructs. A sample was prepared with 20\,v/v$\%$ 2\,$\mu$M DNA constructs in 100 mM NaCl TE buffer, 11\,v/v$\%$ \acp{GUV} (prepared as detailed in the previous paragraph) in 300\,mM glucose solution, and 69\,v/v$\%$ ``correction buffer''. This correction buffer consists of 116\,mM NaCl, 78.2\,mM glucose and 1.16$\times$ \Ac{TE} buffer, and was included to match the osmolarity of the interior and exterior of the \acp{GUV} and thus to prevent osmotic-shock-induced rupture. The sample was left to incubate on a roller mixer overnight. After the functionalisation process, the vesicle sample was washed to minimise the amount of DNA constructs remaining in solution that have not been incorporated into the lipid membranes. This was achieved by leaving the sample to stand for at least 15 minutes to allow the vesicles to settle, and then replacing approximately 90$\%$ of the eluent with an iso-osmolar solution consisting of 100\,mM NaCl and 87\,mM glucose in TE buffer. The washing process was carried out at least three times. To check that the vesicles had been successfully functionalised and the lipid membranes were saturated with DNA constructs, the vesicles were prepared with fluorescently-tagged lipids and DNA constructs, and imaged \emph{via} epifluorescence and confocal microscopy. The estimated density of DNA constructs in the lipid membrane is approximately 0.008\,nm$^{-2}$.

\paragraph{Ligand-density gradient generation and characterisation.}
A 0.5\,$\mu$l droplet of 2\,$\mu$M biotinylated DNA ``ligand'' constructs in buffer solution (100\,mM \ac{NaCl} and 1$\times$\Ac{TE}) was deposited on the substrate, a well plate pre-coated with streptavidin. This was left to incubate for five minutes to allow the DNA constructs to bind to the substrate. Excess DNA was then washed away by quickly rinsing the substrate with buffer solution (100\,mM \ac{NaCl} and 1$\times$\Ac{TE}). This left a coating of DNA in the circular region that had been covered by the droplet, with a diameter of $\approx$3.5\,mm and a ligand density of $\approx$0.033\,nm$^{-2}$. Due to the fast rate of binding between streptavidin and biotin, a small amount of excess DNA bound to the surrounding region during the washing process; the surface density of DNA in this region was measured to be 0.005--0.01\,nm$^{-2}$ on average. At the perimeter of the circular maximally-coated region, there was a steep surface density gradient spanning a distance of approximately 300\,$\mu$m. The ligand constructs were tagged with Alexa Fluor 488, enabling the ligand coverage to be imaged and characterised by excitation of the fluorophore. The ligand density in the maximum-density region of the wells was estimated by applying a droplet of ligand solution to the well for five minutes, like in the gradient generation procedure; the supernatant was diluted and its fluorescence was recorded on a BMG LABTECH CLARIOstar Plus microplate reader. The fluorescence of the diluted supernatant was compared with the fluorescence of ligand solutions of known concentrations, to estimate the quantity of ligand constructs that had been deposited on the substrate. To characterise the ligand-density gradient profiles in experiments, the ligand coverage was imaged with epifluorescence microscopy. Due to non-uniform illumination of the sample, the fluorescence images were normalised by a control image taken within a non-functionalised well containing calcein solution (see Fig.~S6). The ligand density was then estimated by interpolating between the background fluorescence signal (recorded from a non-functionalised well) and the average fluorescence signal of a maximum ligand-density region of the well.

\paragraph{Recording vesicle motion.}
After generating the ligand-density gradients, the receptor-coated vesicles were added to the wells, and the well plate was covered with a transparent adhesive foil to minimise evaporation and evaporation-driven flows in the sample. The sample was then left to sit for one hour to allow most of the vesicles to sink to the bottom of the well and adhere. A time lapse recording was then taken of the vesicles on the substrate, over a duration of 10--15 hours.

\paragraph{Vesicle tracking}
The time lapse recordings were initially stabilised using the \textit{Image Stabilizer} plugin on ImageJ~\cite{ImageStabilizer} to remove any unwanted effects of stage drift or inaccuracies in stage positioning. The image data were then loaded into Python, and processed to remove noise and prepare the images for the tracking algorithm. Vesicles were located in each frame of the time lapse recordings using a template-matching method. Particle trajectories were then identified by linking particle positions between successive time frames; this was achieved using the Python module \textit{Laptrack}~\cite{fukai_2022,laptrack}. More details on data processing and the vesicle locating and tracking algorithms can be found in Section~I of the~SI.

\begin{acknowledgments}
H.S. acknowledges funding from an EPSRC DTP Studentship in the Department of Chemistry at Imperial College London (EP/R513052/1). L.D.M. acknowledges support from the European Research Council (ERC) under the Horizon 2020 Research and Innovation Programme (ERC-STG No 851667 – NANOCELL) and a Royal Society University Research Fellowship (UF160152, URF{\textbackslash}R{\textbackslash}221009). B.M.M.\ is supported by a PDR grant of the FRS-FNRS (Grant No.\ T.0158.21). Y.E. is supported by a UK Research and Innovation Future Leaders Fellowship (MR/S031537/1).
\end{acknowledgments}

\bibliography{main_biblio} 

\end{document}


\newpage
\onecolumngrid

\section{Supplementary data processing methods}
\label{sec:data_processing}
\subsection{Pre-processing vesicle time lapse data}
The time lapse recordings of vesicles over DNA surface density gradients were originally stabilised using the \textit{Image Stabilizer} plugin on ImageJ~\cite{ImageStabilizer} to remove any unwanted effects of stage drift or inaccuracies in stage positioning. The image data were then loaded into Python, and processed to remove noise and prepare the images for the tracking algorithm. The stages of image processing were as follows:

\begin{itemize}
    \item A top-hat transform was applied to the images, to flatten the non-uniform fluorescence signal from the DNA coating on the substrate.
    \item Gaussian smoothing was applied to the images, to remove some fluorescence measurement noise.
    \item The images were binarised, to discern the fluorescent vesicles from the background signal.
    \item Morphological opening operations were applied to the images, to remove noise (\emph{i.e.} bright specks) from the dark background signal.
    \item Morphological closing operations were applied to the images, to remove noise (\emph{i.e.} dark specks) from the bright vesicles.
\end{itemize} 

\subsection{Identifying vesicle trajectories}
Vesicles were located in each frame of the time lapse recordings using a template-matching method. The stages of this method were as follows:

\begin{itemize}
    \item As the vesicle samples are polydisperse, direct application of a template matching method to search for bright circular objects in the binarised images did not perform sufficiently well. For instance, the algorithm would sometimes incorrectly identify a larger vesicle as comprising of multiple smaller vesicles. Therefore, in order to achieve size-invariant identification of vesicles in the images, the distance transform was used. The distance transform was applied to the binarised image data, labelling each pixel with the distance to the nearest dark pixel. 
    \item The distance transform of the vesicle template (also referred to as the kernel) was generated. This template provides the basic shape which the template-matching algorithm searches for in the images.
    \item A template-matching algorithm (the \textit{matchTemplate} functionality of the Python OpenCV module) was applied to the image data, identifying regions in the images which closely match the template (\emph{i.e.} bright disks in the original image). This generated an array of values indicating the similarity of each region of the image to the template.
    \item This array of values was then binarised, to select regions of the image with the highest similarity to the template, which should correspond to the vesicles in the image.
    \item The position and value of the local maxima within each non-contiguous region of high similarity were determined, to provide an initial approximation of the vesicles' positions and radii.
    \item To improve the position and radius estimates to sub-pixel accuracy, a function-fitting approach was employed. We defined a function that outputted the distance transform of a bright disk, depending on the position and size of the disk. We then fitted this function to the local regions of the image in which a vesicle had been approximately located. The fitted model parameters provide a new, more accurate estimate of the vesicle positions.
    \item To try to reduce any remaining noise or inaccuracies, we removed vesicles with an estimated radius of less than one pixel and vesicles overlapping with the edges of the image from the data.
    \item The data was filtered to only include vesicles in the relevant regions of the images \emph{i.e.} in the uniform-density regions or the density-gradient regions. For the former, we manually cropped the data to regions of uniform density; for the latter, we applied a script that identified regions of the image in which there was a steep gradient of DNA density, and then manually cropped out any remaining data in regions that had been incorrectly selected.
\end{itemize}

Vesicle trajectories were then identified by linking vesicle positions between successive time frames; this was achieved using the Python module \textit{Laptrack}~\cite{fukai_2022,laptrack}. This module employs a linear-assignment-problem-based approach to identifying vesicle trajectories.

\subsection{Identifying components of vesicle motion parallel and perpendicular to gradient}
\label{sec:dataprocess_par_perp}
The fluorescence image of the DNA coating on the substrate (prior to addition of the vesicles) was processed as follows:

\begin{itemize}
    \item Gaussian smoothing was applied to the image to reduce noise in the fluorescence signal.
    \item To account for any uneven illumination or camera sensitivity across the frame of view, the image was normalised by a control image of a uniform fluorescence signal. This control image was attained by imaging a well filled with calcein solution.
\end{itemize}

\noindent The vector gradient (also referred to as the directional derivative) of the fluorescence image of the DNA-coated substrate was computed. The resulting vector field describes the magnitude and direction of the density gradient at each point in the image. Finally, we calculated the components of the vesicle velocities parallel and perpendicular to the direction of the gradient. This was achieved by calculating the vector dot and cross products (respectively) of the vesicle velocities with the unit vectors describing the direction of the gradient at each vesicle position.

\FloatBarrier
\subsection{Mean squared displacement analysis and filtering of vesicle data}
\label{subsec:msd}
We employed \ac{MSD} analysis to study vesicle motion within our experimental data. For each vesicle trajectory, we calculated the \ac{MSD} for a given lag time $\tau$ by slicing the trajectory into sections of duration $\tau$ and averaging the square displacement over these sections, as follows:
\FloatBarrier

\begin{equation}
MSD(\tau) = \frac{1}{\lfloor \frac{T}{\tau} \rfloor}  \sum_{n=0}^{\lfloor \frac{T}{\tau} \rfloor - 1} (\vec{r}_{(n+1)\tau}-\vec{r}_{n\tau})^2
\end{equation}

\noindent where $T$ is the total recorded duration of the trajectory, and $\vec{r}_{t}$ is the position of the vesicle at time $t$.

We then plotted the \ac{MSD} as a function of the lag time, to identify the type of motion. In general, particles undergoing Brownian diffusion exhibit a linear relationship between \ac{MSD} and lag time, while particles travelling with directed motion exhibit a quadratic relationship.\cite{einstein1905molekularkinetischen,von1906kinetischen}

\begin{figure}
    \centering
    \includegraphics{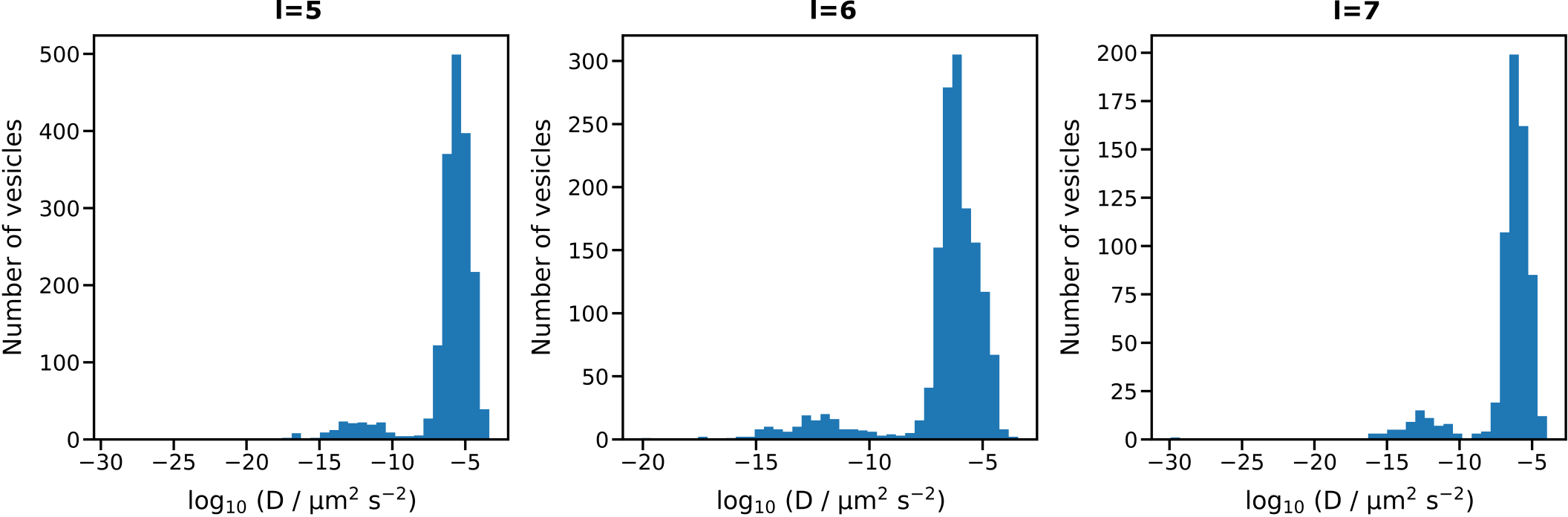}
    \caption{\textbf{Estimated diffusion coefficients of vesicles measured in experiments.} The histograms illustrate the distribution of diffusion coefficients, $D$, estimated for experimental systems with sticky end length $l=$~5, 6 and 7\,nt. The diffusion coefficients were estimated from experimental trajectory data \emph{via} the method outlined in Section~\ref{subsec:msd} of the SI. By plotting the decimal logarithm of the vesicle diffusion coefficients on the x-axis, bimodal distributions of the estimated diffusion coefficients are visible.}
    \label{fig:bimodal_diffconstant}
\end{figure}

To gain an initial insight into our experimental data, we used a least-squares approach to fit straight lines to the plots of \ac{MSD} versus lag time. The straight lines were only fitted to values of \ac{MSD} that had been averaged over at least two trajectory slices, and non-negative constraints were enforced on the intercept and slope of the fitted lines, as negative values would be non-physical and indicate tracking errors. We then extracted rough estimates of the vesicle diffusion constants from the slope of the line. Upon plotting histograms of the estimated diffusion constants, we observed bimodal distributions as seen in Figure \ref{fig:bimodal_diffconstant}, with two distinct peaks corresponding to a ``faster" and ``slower" vesicle population. For each of the $l=$~5, 6 and 7\,nt sticky end systems, approximately 5--6$\%$ of the vesicles appear to belong to the ``slower" population.

To understand the underlying factor behind the bimodal distribution, we inspected the \ac{MSD} plots as a function of lag time for vesicles in the ``slower" and ``faster" populations of the bimodal distribution. We found that for the majority of the vesicles in the ``slower" population, the \ac{MSD} appeared to have a decreasing trend with increasing lag time. Due to the non-negative constraints on the intercept and slope of the linear fits, the fitting algorithm assigned negligible, near-zero values to the fitted parameters, resulting in an apparent secondary population of slower-moving vesicles.

We then studied the trajectories of the vesicles with an apparent negative relationship between \ac{MSD} and lag time, by comparing the algorithmically tracked trajectories with the imaged vesicle motion by eye. In the majority of these cases it appeared that inaccuracies and noise in the particle tracking had resulted in the negative relationship between \ac{MSD} and lag time.

We also observed by eye that a small minority of vesicles in our experiments appear to be immobile. This could in part be due to non-uniformity in DNA density on the substrate and vesicle surfaces resulting in vesicles being trapped in high density regions. However, we still observed a small proportion of immobile vesicles in systems where no DNA ligands or receptors were present. We believe that these immobile vesicles may be become pinned to surface defects, or trapped by strong non-specific adhesion. Bimodal distributions of particle motility with an immobile population have also been observed in other particle motility studies~\cite{hamming2023receptor}.

Finally, we note that erroneous tracking of sample impurities or surface defects as vesicles could contribute to the apparent population of immobile vesicles.

We chose to discard the apparent ``slow" population of vesicles from the experimental data from further data analysis, by excluding vesicles with a diffusion constant estimated to be less than 10$^{-8}$ ${\mu}$m$^{2}$s$^{-2}$. This would reduce the number of inaccurate trajectories included in our data analysis, and also the number of vesicles that are immobile for non-DNA reasons. However, for transparency, we present our results pre- and post-filtering in the SI.
\FloatBarrier

\newpage
\section{Supplementary numerical methods}
\subsection{Evaluation of timescales relevant to vesicle dynamics}
\label{subsec:timescales}

\subsubsection{Reconfiguration timescale of constraining bridges}

The translational configurational space available to the vesicle depends on a set of constraining bridges near the perimeter of the contact region. The timescale of the reconfiguration of these constraining bridges can be estimated as $\tau_\mathrm{r}=1/(2 k_\mathrm{off} n_\mathrm{cb})$, where $k_\mathrm{off}$ is the rate at which bridges break, and with the average number of constraining bridges $\langle n_\mathrm{cb}\rangle \approx5$.\cite{lowensohn2022sliding} From Eq.~1 in the main text we calculated $k_\mathrm{off}=$~1.10\,s$^{-1}$, 0.174\,s$^{-1}$, and 0.0084\,s$^{-1}$ for sticky ends with $l=
$~5, 6, and 7\,nt respectively, under experimental conditions. Using these values, we estimate the timescales of constraining bridge reconfiguration to be 0.091\,s, 0.57\,s, and 11.9\,s for sticky ends with $l=
$~5, 6, and 7\,nt respectively.

\subsubsection{Diffusion timescale}
\begin{figure}[H]
    \centering
    \includegraphics{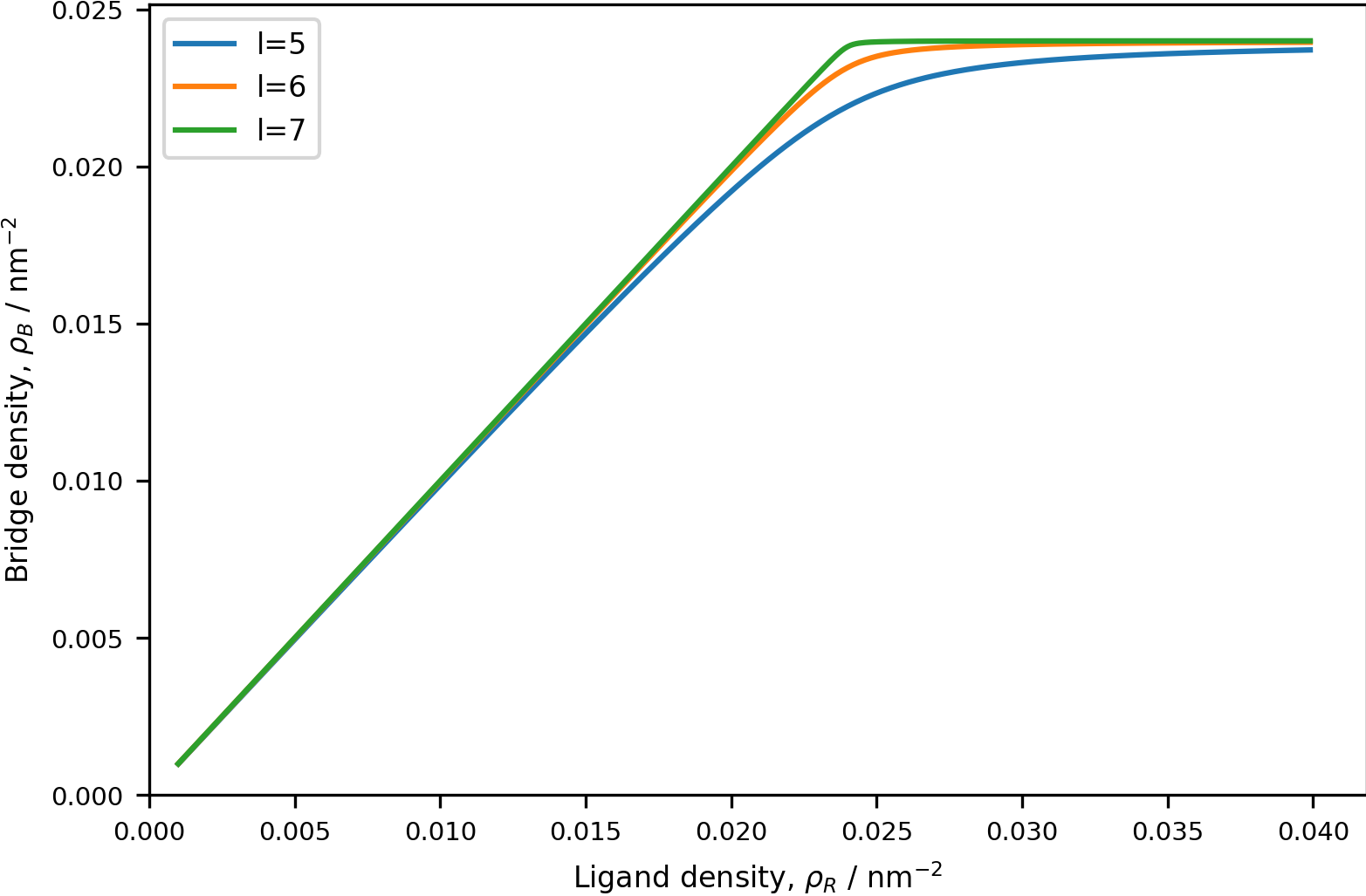}
    \caption{\textbf{Under experimental conditions, the density of receptor-ligand bridges is saturated and reaches its maximal value.} The bridge density $\rho_B$ is plotted against ligand density $\rho_L$, while the initial density of receptors over the vesicle membrane $\rho_R$ is fixed at 0.008\,nm$^{-2}$, for 0.2\,$\mu$m diameter vesicles and sticky ends lengths $l=$~5, 6 and 7\,nt (DNA sequences shown in Table~\ref{tab:strandsequences1}). The bridge density has been estimated using Eq.~\ref{Eq:rhoB}, with the equilibrium constant $K_{eq}$ estimated using the Arrhenius equation $K_{eq}=\frac{e^{-\beta\Delta_G}}{\rho_0 LA_{tot}}$ where $\Delta_G$ is the standard hybridisation energy of the sticky ends, $\rho_0=$~1\,M is the standard concentration, $L$ is the length of the sticky ends and $A_{tot}$ is the total surface area of the vesicle. While $\rho_L<3\rho_R$ the bridge density is limited by $\rho_L$, whereas when $\rho_L>3\rho_R$ the bridge density saturates at $3\rho_L$. The factor of 3 results from the redistribution of the receptors from an initial uniform concentration across the vesicle (approximated as a hemisphere) to being concentrated within the contact region (a third of the total surface area of the hemisphere).}
    \label{fig:bridge_saturation}
\end{figure}

The diffusion timescale needed by a vesicle to explore the available configurational space can be estimated as $\tau_\mathrm{d}=\pi R^2 n_\mathrm{cb} / (D n_\mathrm{b}^2)$, where $D$ is the diffusion constant of free (non-adhering) vesicles. The Stokes-Einstein equation can be used to estimate the diffusion constant of a 5\,$\mu$m-radius vesicle as 0.049\,$\mu$m$^2$s$^-1$. In Fig.~\ref{fig:bridge_saturation} we note that the bridge density is comparable to the ligand and density receptors, \emph{i.e.} on the order of 0.01\,nm$^{-2}$. Using these values, we estimate the timescale for the vesicle to explore the available configurational space via diffusion to be 1.30$\times$10$^{-8}$\,s.

\subsection{Analytic prediction of vesicle drifting velocity}
\label{Sec:driftvel}
In this section, we derive an expression of the drifting velocity ($v^{FD}_x$) using our recent estimation of the diffusion constant $D(x_\mathrm{CM})$ of particles with mobile ligands \cite{lowensohn2022sliding} 
\begin{eqnarray}
D(x_\mathrm{CM})=c \frac{k_\mathrm{off}}{\langle n_b(x_\mathrm{CM})\rangle ^2 } , 
\end{eqnarray}
where $x_\mathrm{CM}$ is the particle's center of mass, $n_b$ the number of bridges, and $c=\pi R^2/2$ in the weak-binding regime ($n_b\ll N_L$, where $N_L$ is the number of ligands underneath the vesicle) with the vesicle radius $R$. In particular, we assume the following expression for the drifting velocity 
\begin{eqnarray}
v^{FD}_x(x_\mathrm{CM}) = - \beta D (x_\mathrm{CM})  F'(x_\mathrm{CM})  
\end{eqnarray}
following from the fluctuation-dissipation theorem \cite{callen1951irreversibility}, where $f'(x_\mathrm{CM})=\mathrm{d}f(x_\mathrm{CM})/\mathrm{d} x_\mathrm{CM}$ and 
$F(x_\mathrm{CM})$ is the free energy of the system at a given $x_\mathrm{CM}$ \cite{mognetti2019programmable}
\begin{eqnarray}
\beta F(x_\mathrm{CM}) = N_R \log\left[1 - \frac{n_b(x_\mathrm{CM})}{N_R}\right] + N_L(x_\mathrm{CM}) \log \left[1 - \frac{n_b(x_\mathrm{CM})}{N_L(x_\mathrm{CM})}\right]+n_b(x_\mathrm{CM}) \, ,
\label{Eq:F}
\end{eqnarray}
where $N_R$ is the total number of receptors over the vesicle membrane. We have (see Sec.~\ref{Sec:MP})
\begin{eqnarray}
- \beta \frac{\mathrm{d} F(x_\mathrm{CM})}{\mathrm{d} x_\mathrm{CM}}
=-N_L'(x_\mathrm{CM})  \log\left[ 1- \frac{n_b(x_\mathrm{CM})}{N_L (x_\mathrm{CM}) } \right]
\label{Eq:DriftingForce}
\end{eqnarray}
from which we derive 
\begin{eqnarray}
v^{FD}_x(x_\mathrm{CM})  &=& -\frac{\pi R^2 k_\mathrm{off}}{2 n_b(x_\mathrm{CM})^2} N_L(x_\mathrm{CM})' \log\left[ 1- \frac{n_b(x_\mathrm{CM})}{N_L (x_\mathrm{CM}) } \right] 
\end{eqnarray}
For a linear gradient of ligands ($\rho(x)=c_0 + \lambda x$, if $\rho$ is the ligands' density) we have $N_L(x_\mathrm{CM})'=\lambda \pi R^2$. Therefore 
\begin{eqnarray}
v^{FD}_x(x_\mathrm{CM})  &=& - \frac{\lambda k_\mathrm{off}}{2} \left[\frac{\pi R^2}{n_b(x_\mathrm{CM})  }\right]^2 
\log\left[ 1- \frac{n_b(x_\mathrm{CM})}{N_L(x_\mathrm{CM})} \right]
\\
&=& - \frac{ \lambda k_\mathrm{off}}{2} \left[ \frac{\pi R^2}{n_b(x_\mathrm{CM})} \right]^2
\log\left[ 1- \frac{n_b(x_\mathrm{CM})}{\pi R^2 \rho(x_\mathrm{CM}) } \right]
\\
&=& - \frac{\lambda k_\mathrm{off}}{2\rho_b (x_\mathrm{CM})^2 }  \log\left[1- \frac{ \rho_b(x_\mathrm{CM})}{\rho(x_\mathrm{CM}) } \right]
\label{Eq:vdrift_lin}
\end{eqnarray}
where $\rho_b(x_\mathrm{CM})$ is the density of bridges at the particle's center of mass and $\rho$ is the density of ligands. 
The last equality in the previous equation follows from the fact that, for a given $x_\mathrm{CM}$, $\rho_b(x)$ is proportional to $\rho(x)$ (see Eq.~\ref{Eq:rhoB}).
In the weak binding limit [$\rho_b(x_\mathrm{CM}) \ll \rho(x_\mathrm{CM})$], $v^{FD}_x$ further simplifies as follows 
\begin{eqnarray}
v^{FD}_x (x_\mathrm{CM}) =
\frac{\lambda k_\mathrm{off}}{2\rho_b (x_\mathrm{CM}) \rho (x_\mathrm{CM}) } \, .  
\end{eqnarray}
Note that the weak-binding regime is not relevant to the experimental conditions used in this work.

As the diffusion constant ($D$) is not homogeneous, it follows that $v^{FD}_x(x_\mathrm{CM})$ may include an extra term ($v^{FD}_{x,\alpha}$) equal to \cite{sokolov2010ito}
\begin{eqnarray}
v^{FD}_{x,\alpha}(x_\mathrm{CM}) = D'(x_\mathrm{CM}) = - 2\frac{\pi R^2 k_\mathrm{off}}{n_b(x_\mathrm{CM})^3 } n_b'(x_\mathrm{CM}) \, .
\end{eqnarray}
Using Eq.~\ref{Eq:nBp} to express $n_b'$ in terms of $N_L'$, we obtain
\begin{eqnarray}
v^{FD}_{x,\alpha}(x_\mathrm{CM})  &=& - 2 \lambda k_\mathrm{off} \left[\frac{\pi R^2}{n_b(x_\mathrm{CM})  }\right]^2 
 \frac{N_R - n_b(x_\mathrm{CM})}{N_L(x_\mathrm{CM}) N_R - n_b(x_\mathrm{CM})^2 }
\\
&=& - 2 \frac{\lambda k_\mathrm{off}}{\rho_b (x_\mathrm{CM})^2 }  \frac{\rho_L - \rho_b(x_\mathrm{CM})}{\rho(x_\mathrm{CM}) N_R - \rho_b(x_\mathrm{CM}) n_b(x_\mathrm{CM}) }
\end{eqnarray}
where $\rho_L$ is the density of ligands. 
$v^{FD}_{x,\alpha}(x_\mathrm{CM})$ is expected to be negligible as it divides an intensive quantity by an extensive one.

\subsection{Proof of Eq.~\ref{Eq:DriftingForce}}\label{Sec:MP}

\subsubsection{Number and density of bridges} 

We first calculate the equilibrium predictions for the density and number of bridges ($\rho_b (x_\mathrm{CM})$ and $n_b(x_\mathrm{CM})$, respectively). The density of bridges underneath the particles at a given position $x$ along the gradient solves the following equation 
\begin{eqnarray}
\rho_b(x) &=& K_\mathrm{eq} \left[ \rho(x)-\rho_b(x) \right] \left[ N_R-n_b(x_\mathrm{CM}) \right]
\label{Eq:rho_B}
\end{eqnarray}
where $K_\mathrm{eq}$ is the equilibrium constant controlling the formation of bridges \cite{mognetti2019programmable}. The number of bridges and the total number of ligands underneath the particles are obtained by integrating, respectively, $\rho_b(x)$ and $\rho(x)$ over the contact region between the vesicle and the surface ($\Omega(x_\mathrm{CM})$)
\begin{eqnarray}
n_b(x_\mathrm{CM}) = \int_{\Omega(x_\mathrm{CM})} \rho_b(x) \mathrm{d} x \mathrm{d} y
&\qquad&
N_L(x_\mathrm{CM}) = 
\int_{\Omega(x_\mathrm{CM})} \rho(x) \mathrm{d} x \mathrm{d} y
\end{eqnarray}
where $y$ is the coordinate orthogonal to the gradient. By integrating Eq.~\ref{Eq:rho_B} over $\Omega(x_\mathrm{CM})$ we find 
\begin{eqnarray}
n_b(x_\mathrm{CM}) &=& K_\mathrm{eq} \left[N_L(x_\mathrm{CM})-n_b(x_\mathrm{CM})  \right]
\left[ N_R-n_b(x_\mathrm{CM}) \right]
\label{Eq:nB}
\end{eqnarray}
which is solved as follows   
\begin{eqnarray}
n_b(x_\mathrm{CM}) &=& \frac{ K_\mathrm{eq} N_L(x_\mathrm{CM}) + K_\mathrm{eq} N_R + 1 - \sqrt{\Delta^2}}{2 K_\mathrm{eq}}
\\
\Delta^2 &=&  \left[ K_\mathrm{eq} N_L(x_\mathrm{CM}) - K_\mathrm{eq} N_R \right]^2 + 2 K_\mathrm{eq} N_L(x_\mathrm{CM}) + 2 K_\mathrm{eq} N_R + 1
\end{eqnarray} 
Given $n_b(x_\mathrm{CM})$, the density of bridges reads as follows 
\begin{eqnarray}
\rho_b(x) &=& \frac{K_\mathrm{eq} \left[ N_R-n_b(x_\mathrm{CM}) \right]}{1 + K_\mathrm{eq} \left[ N_R-n_b(x_\mathrm{CM}) \right] } \rho (x)
\label{Eq:rhoB}
\end{eqnarray}

\subsubsection{Drifting force} 

We now prove Eq.~\ref{Eq:DriftingForce} using Eq.~\ref{Eq:nB}. From Eq.~\ref{Eq:F} we calculate
\begin{eqnarray}
\beta F'(x_\mathrm{CM}) &=& - \frac{n_B'(x_\mathrm{CM}) N_R}{N_R - n_B(x_\mathrm{CM})} + \frac{n_B(x_\mathrm{CM}) N_L'(x_\mathrm{CM})-N_L(x_\mathrm{CM})n_B'(x_\mathrm{CM})}{N_L(x_\mathrm{CM})-n_B(x_\mathrm{CM})}+n_B'(x_\mathrm{CM})
\nonumber \\
&& +N_L'(x_\mathrm{CM})\log\left[1- \frac{n_B(x_\mathrm{CM})}{N_L(x_\mathrm{CM})} \right]
\label{Eq:Fp}
\\
& \equiv &
\Lambda + N_L'(x_\mathrm{CM})\log\left[1- \frac{n_B(x_\mathrm{CM})}{N_L(x_\mathrm{CM})} \right]
\nonumber
\end{eqnarray}
By deriving Eq.~\ref{Eq:nB} we obtain 
\begin{eqnarray}
n_B'(x_\mathrm{CM}) &=& K_\mathrm{eq} \left[N_L'(x_\mathrm{CM})-n_B'(x_\mathrm{CM})  \right]
\left[ N_R-n_B(x_\mathrm{CM}) \right] - K_\mathrm{eq} \left[N_L(x_\mathrm{CM})-n_B(x_\mathrm{CM})  \right] n_B'(x_\mathrm{CM})
\nonumber \\
&=& \frac{n_B(x_\mathrm{CM}) \left[N_L'(x_\mathrm{CM})-n_B'(x_\mathrm{CM})\right]}{N_L(x_\mathrm{CM}) - n_B(x_\mathrm{CM}) } - \frac{ n_B(x_\mathrm{CM}) n_B'(x_\mathrm{CM})}{N_R - n_B (x_\mathrm{CM})}
\end{eqnarray}
where in the second line we have used Eq.~\ref{Eq:nB}. From the previous equation we obtain an expression for $n_B'(x_\mathrm{CM})$ as a function of $N_L'(x_\mathrm{CM})$
\begin{eqnarray}
n_B'(x_\mathrm{CM}) &=& \frac{\left[N_R - n_B(x_\mathrm{CM}) \right] n_B(x_\mathrm{CM}) N_L'(x_\mathrm{CM})}{N_L(x_\mathrm{CM}) N_R - n_B(x_\mathrm{CM})^2} 
\label{Eq:nBp}
\end{eqnarray}
By using the previous equation in Eq.~\ref{Eq:Fp} we can prove that the first three terms in the rhs of the latter cancel, therefore proving Eq.~\ref{Eq:DriftingForce}. Specifically 
\begin{eqnarray}
\Lambda &=& - \frac{n_B'(x_\mathrm{CM}) N_R}{N_R - n_B(x_\mathrm{CM})} + \frac{n_B(x_\mathrm{CM}) N_L'(x_\mathrm{CM})-N_L(x_\mathrm{CM}) n_B'(x_\mathrm{CM})}{N_L(x_\mathrm{CM})-n_B(x_\mathrm{CM})}+n_B'(x_\mathrm{CM})
\nonumber
\\
&=& - \frac{ n_B(x_\mathrm{CM}) N_R N_L'(x_\mathrm{CM})}{N_L(x_\mathrm{CM}) N_R - n_B(x_\mathrm{CM})^2} + \frac{n_B(x_\mathrm{CM}) N_L'(x_\mathrm{CM})}{N_L(x_\mathrm{CM})-n_B(x_\mathrm{CM})} 
\nonumber\\
&&- \frac{ N_L(x_\mathrm{CM})}{N_L(x_\mathrm{CM})-n_B(x_\mathrm{CM})} \frac{\left[N_R - n_B(x_\mathrm{CM})\right] n_B(x_\mathrm{CM}) N_L'(x_\mathrm{CM})}{N_L(x_\mathrm{CM}) N_R - n_B(x_\mathrm{CM})^2} 
\nonumber\\
&&+\frac{\left[N_R - n_B(x_\mathrm{CM})\right] n_B(x_\mathrm{CM}) N_L'(x_\mathrm{CM})}{N_L(x_\mathrm{CM}) N_R - n_B(x_\mathrm{CM})^2}
\nonumber
\\
&=&  \frac{n_B(x_\mathrm{CM}) N_L'(x_\mathrm{CM})}{N_L(x_\mathrm{CM})-n_B(x_\mathrm{CM})} - \frac{ N_L(x_\mathrm{CM})}{N_L(x_\mathrm{CM})-n_B(x_\mathrm{CM})} \frac{\left[N_R - n_B(x_\mathrm{CM})\right] n_B(x_\mathrm{CM}) N_L'(x_\mathrm{CM})}{N_L(x_\mathrm{CM}) N_R - n_B(x_\mathrm{CM})^2} 
\nonumber\\
&&- \frac{ n_B(x_\mathrm{CM})^2 N_L'(x_\mathrm{CM})}{N_L(x_\mathrm{CM}) N_R - n_B(x_\mathrm{CM})^2}
\nonumber
\\
&=& \frac{n_B(x_\mathrm{CM}) N_L'(x_\mathrm{CM})}{N_L(x_\mathrm{CM})-n_B(x_\mathrm{CM})}
\nonumber\\
&&-\frac{n_B(x_\mathrm{CM}) N_L'(x_\mathrm{CM})}{N_L(x_\mathrm{CM}) N_R - n_B(x_\mathrm{CM})^2}
\left[
\frac{ N_L(x_\mathrm{CM}) \left[N_R-n_B(x_\mathrm{CM})\right]}{N_L(x_\mathrm{CM})-n_B(x_\mathrm{CM})} + n_B(x_\mathrm{CM})
\right]
\nonumber
\\
&=& \frac{n_B(x_\mathrm{CM}) N_L'(x_\mathrm{CM})}{N_L(x_\mathrm{CM})-n_B(x_\mathrm{CM})} -  \frac{n_B(x_\mathrm{CM}) N_L'(x_\mathrm{CM})}{N_L(x_\mathrm{CM}) N_R - n_B(x_\mathrm{CM})^2} \frac{N_L(x_\mathrm{CM}) N_R - n_B(x_\mathrm{CM})^2}{N_L(x_\mathrm{CM}) - n_B(x_\mathrm{CM})}
\nonumber
\\
&=& 0
\nonumber
\end{eqnarray}

\newpage
\FloatBarrier
\section{Supplementary tables}
\begin{table}[h]
    \caption{Sequences of all DNA strands used in this work (5' to 3'). The substrate-anchored ligand constructs were manufactured from equal parts of strands S1, S5 and S6, while the membrane-anchored receptor constructs were manufactured from equal parts of strands S2, S3 and S4, with S3. Strands S2 and S5 feature complementary sticky ends, and sequences are displayed for systems with sticky end lengths varying from three to ten bases. The ligand and receptor constructs feature biotin and cholesterol/cholesteryl modifications respectively for anchoring to the relevant surfaces. For fluorescent labelling of the ligands and receptors, strands S1 and S2 are labelled with the fluorophore Alexa Fluor 488. Schematics of the constructs are depicted in Figure \ref{fig:dnastrandsschematics}.}
    \begin{tabular}{|p{0.1\textwidth}|p{0.07\textwidth}|p{0.2\textwidth}|p{0.58\textwidth}|}
        \hline	 
        \textbf{Construct} & \textbf{Strand} & \textbf{Sticky end length\,/\,nt }& \textbf{Sequence (5' to 3')} \\ 
        \hline 
        Ligand & S1 & 3--10 & CGCGACTTCCTCGCCGCG CGCGAGTTCGAGCTACGC (---AF488) \\ 
        \cline{2-4} 
        & S5 & 3 & GCG TT GCGTAGCTCGAACTCGCG \\ 
        
        &  & 4 & GCGG TT GCGTAGCTCGAACTCGCG \\ 
        
        &  & 5 & GCGGC TT GCGTAGCTCGAACTCGCG \\ 
        
        &  & 6 & GCGTGC TT GCGTAGCTCGAACTCGCG \\ 
        
        &  & 7 & CGCACCG TT GCGTAGCTCGAACTCGCG \\ 
        
        &  & 8 & GTGGACGC TT GCGTAGCTCGAACTCGCG \\ 
        
        &  & 9 & GGTCGCAGC TT GCGTAGCTCGAACTCGCG \\ 
        
        &  & 10 & CGTCCGTGCC TT GCGTAGCTCGAACTCGCG \\ 
        \cline{2-4} 
        & S6 & 3--10 & CGCGGCGAGGAAGTCGCG --- Biotin \\ 
        \hline 
        Receptor & S2 & 3 & CGC TT GAGAGTAGGACCGGCGCG \\ 
        
        &  & 4 & CCGC TT GAGAGTAGGACCGGCGCG \\ 
    
        &  & 5 & GCCGC TT GAGAGTAGGACCGGCGCG \\ 
        
        &  & 6 & GCACGC TT GAGAGTAGGACCGGCGCG \\ 
        
        &  & 7 & CGGTGCG TT GAGAGTAGGACCGGCGCG (---AF488) \\ 
        
        &  & 8 & GCGTCCAC TT GAGAGTAGGACCGGCGCG \\ 
        
        &  & 9 & GCTGCGACC TT GAGAGTAGGACCGGCGCG \\ 
        
        &  & 10 & GGCACGGACG TT GAGAGTAGGACCGGCGCG \\ 
        \cline{2-4} 
        & S3 & 3--10 & CGTTTGCAGGAACGAGAC TT --- Cholesterol-TEG \\ 
        \cline{2-4} 
        & S4 & 3--10 & Cholesteryl-TEG --- TT GTCTCGTTCCTGCAAACG CGCGCCGGTCCTACTCTC \\
        \hline 
    \end{tabular} 
    \label{tab:strandsequences1}
\end{table}
\FloatBarrier

\newpage
\FloatBarrier
\section{Supplementary figures}
\begin{figure}[h!]
\centering
    \includegraphics{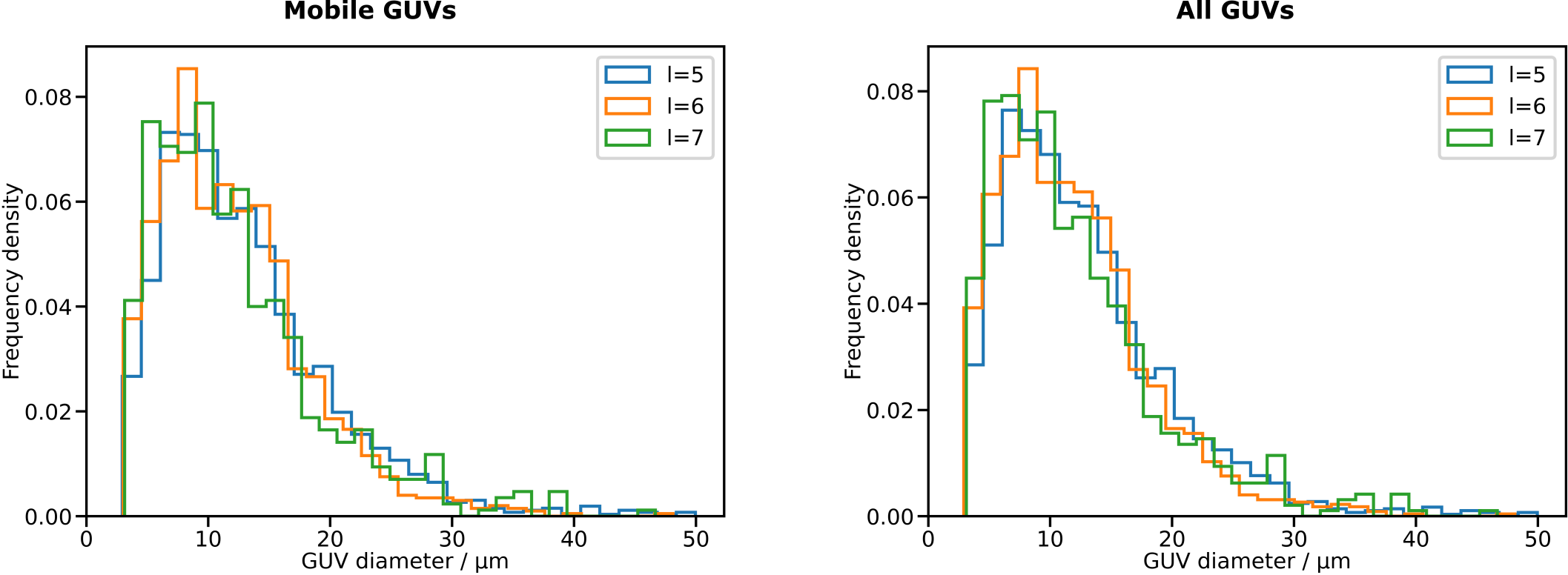}
    \caption{\textbf{Size distributions of vesicles recorded in experiments.} Frequency-density histograms are shown for the filtered (mobile) vesicle population (left) and for all vesicles (right). The filtering method is outlined in Section~\ref{subsec:msd}. The vesicles have been produced \emph{via} electroformation, which generates polydisperse samples with median diameters ranging between 10--11\,$\mu$m and mean diameters ranging between 12--13\,$\mu$m. Note that as part of our vesicle tracking algorithm, we exclude vesicles with diameters measured to be less than 2\,$\mu$m, to avoid erroneous tracking of noise or bright specks in the image. The method of tracking the vesicles and measuring their diameters is further detailed in Section~\ref{sec:data_processing} of the SI.}
    \label{fig:vesicleSizeDist}
\end{figure}

\clearpage

\begin{figure}[h]
	\centering
	\includegraphics[width=10cm]{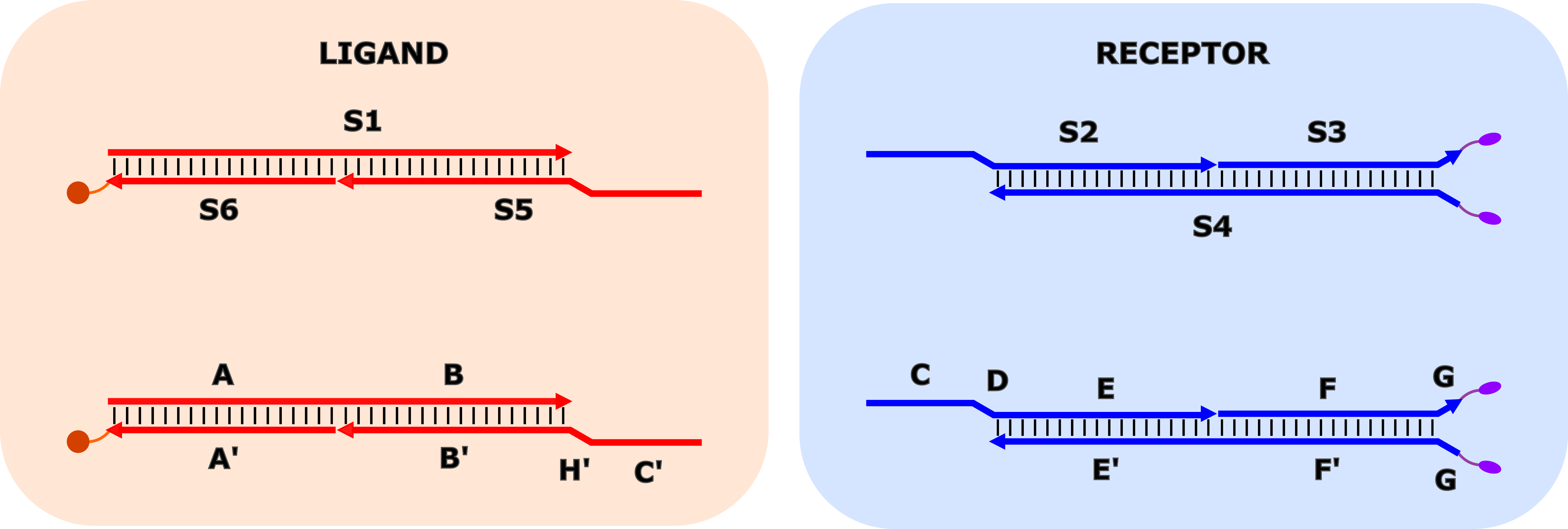}
	\caption{\textbf{Schematics of the DNA ligand and receptor constructs used in this work, labelled by their constituent strands (top) and by their domain structure (bottom).} Domains are separated by white spaces in the corresponding sequences of Table~\ref{tab:strandsequences1}. Domains $C$ and $C'$ are complementary sticky ends.}
	\label{fig:dnastrandsschematics}
\end{figure}
\FloatBarrier

\clearpage

\begin{figure}
    \centering
    \includegraphics{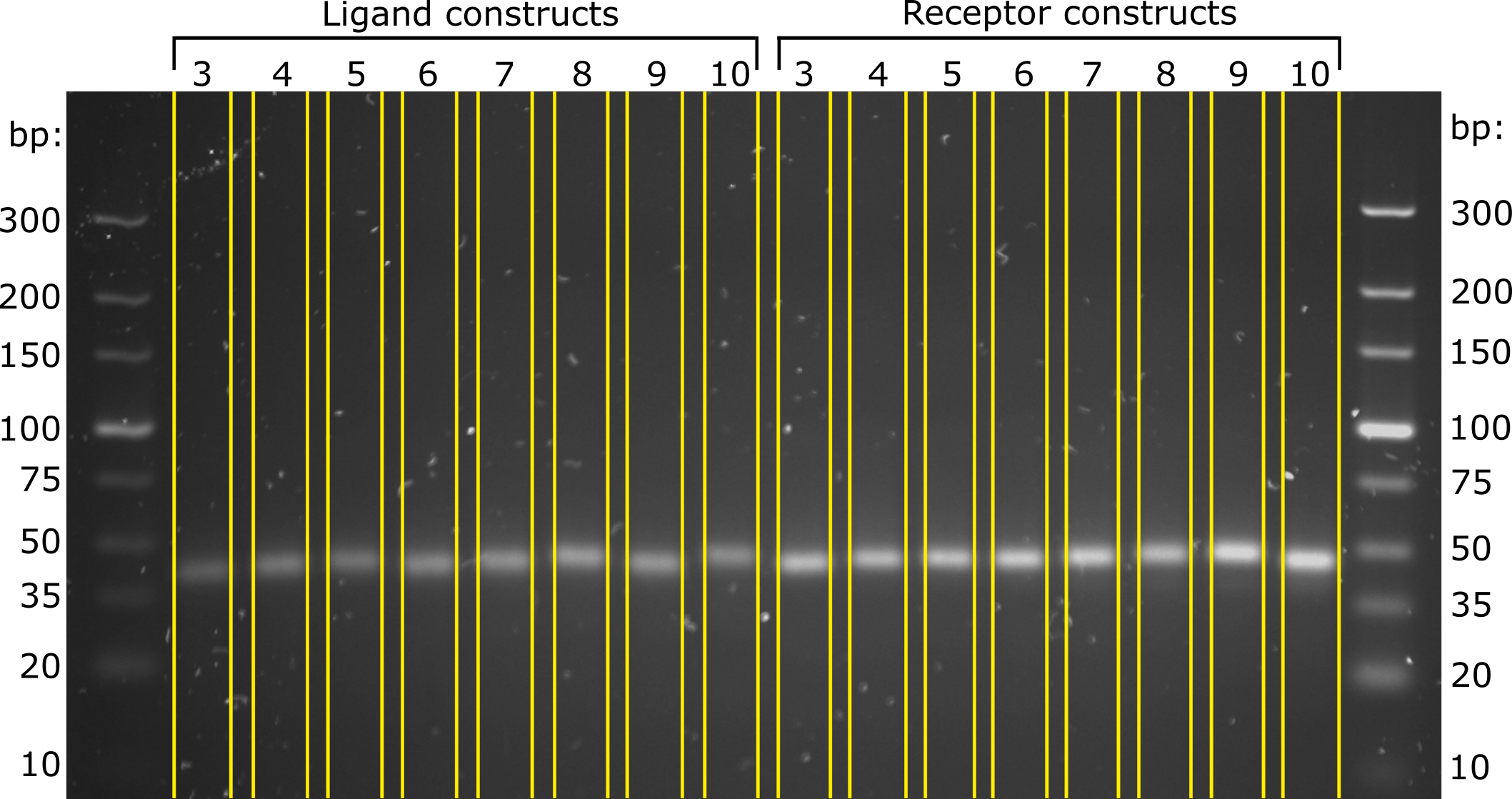}
    \caption{\textbf{Gel electrophoresis results for DNA ligand and receptor constructs.} Results for ligands and receptors with sticky end lengths $l=$~3--10\,nt are shown. A sharp peak between the 35 and 50\,bp reference markers of the DNA ladders is observed for all constructs, which agrees with the expected construct sizes.}
    \label{fig:gelEP_constructs}
\end{figure}

\clearpage

\begin{figure}
    \centering
    \includegraphics{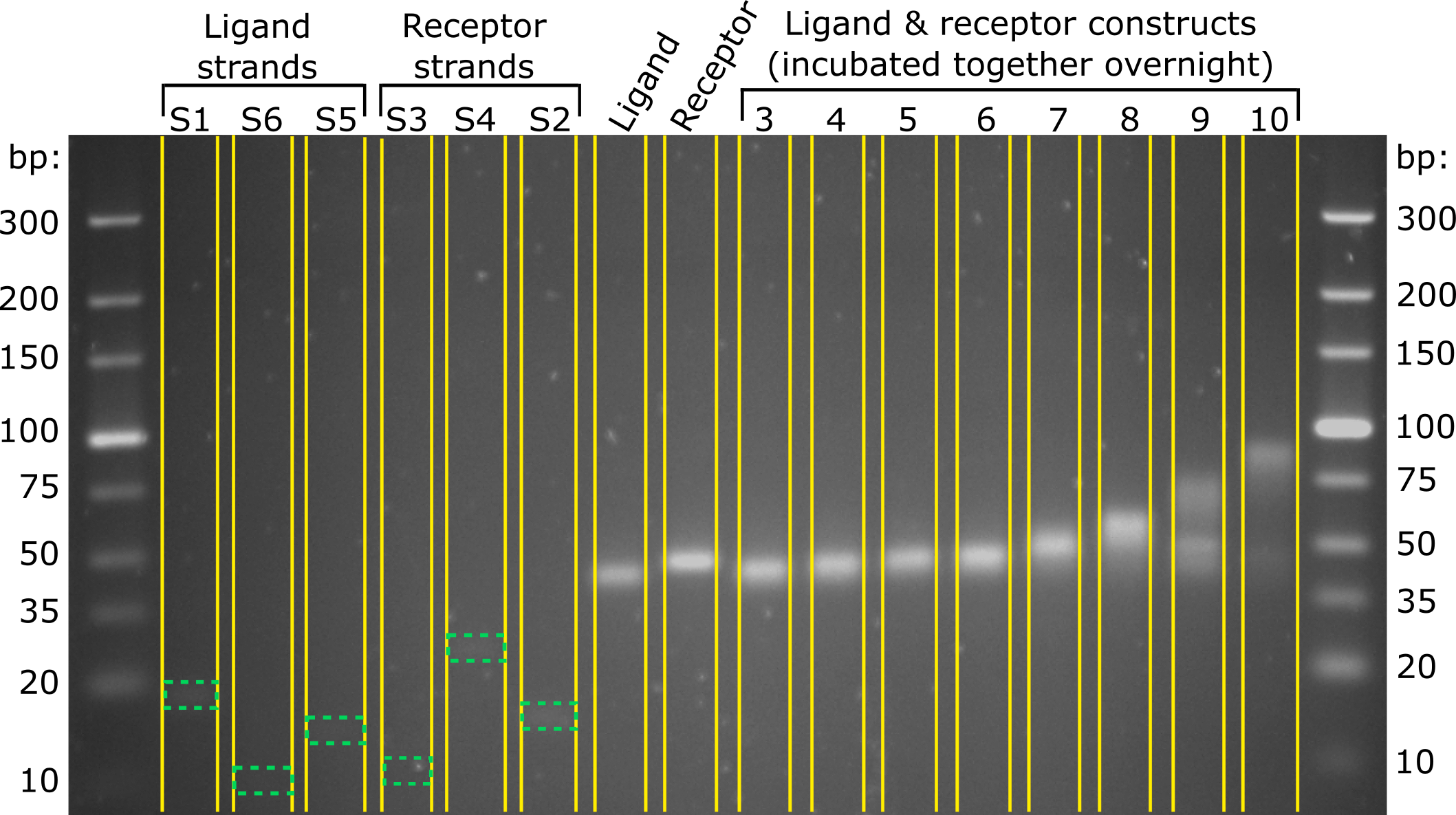}
    \caption{\textbf{Gel electrophoresis results verifying formation of ligand and receptor constructs, and increased binding strength between complementary sticky ends as sticky end length increases.} On the left half of the gel, results are shown for individual ligand and receptor strands (S1--6) and constructs with sticky end lengths $l=$~5\,nt. As the gel stain (SYBR Safe) preferentially binds to double-stranded DNA, the bands for the individual ligand and receptor strands are significantly fainter than the bands for the annealed, double-stranded constructs; the positions of the single-stranded DNA bands have been annotated with dashed boxes to aid with visualisation. On the right half of the gel, results are shown for samples where 1\,$\mu$M ligand and 1\,$\mu$M receptor were incubated together overnight, for $l=$~3--10\,nt; as $l$ increases, the presence of a second, larger species (a ligand bound to a receptor) in the sample becomes more pronounced.}
    \label{fig:gelEP_constructs2}
\end{figure}

\clearpage

\begin{figure}
    \centering
    \includegraphics[width=0.4\columnwidth]{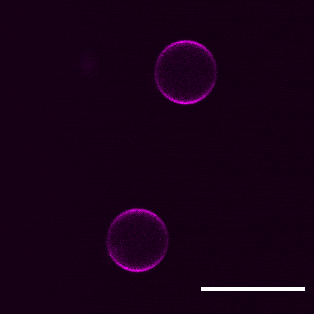}
    \caption{\textbf{Confocal image of two vesicles functionalised with DNA ``receptor'' constructs}. In this case, the fluorescence signal is from the Alexa Fluor 48 tag on the receptor constructs bound to the membrane. The receptors have a sticky end length of 7\,nt. The scale bar indicates a length of 20\,$\mu$m.}
    \label{fig:vesicleFunc}
\end{figure}

\clearpage

\begin{figure}
    \centering
    \includegraphics[width=12cm]{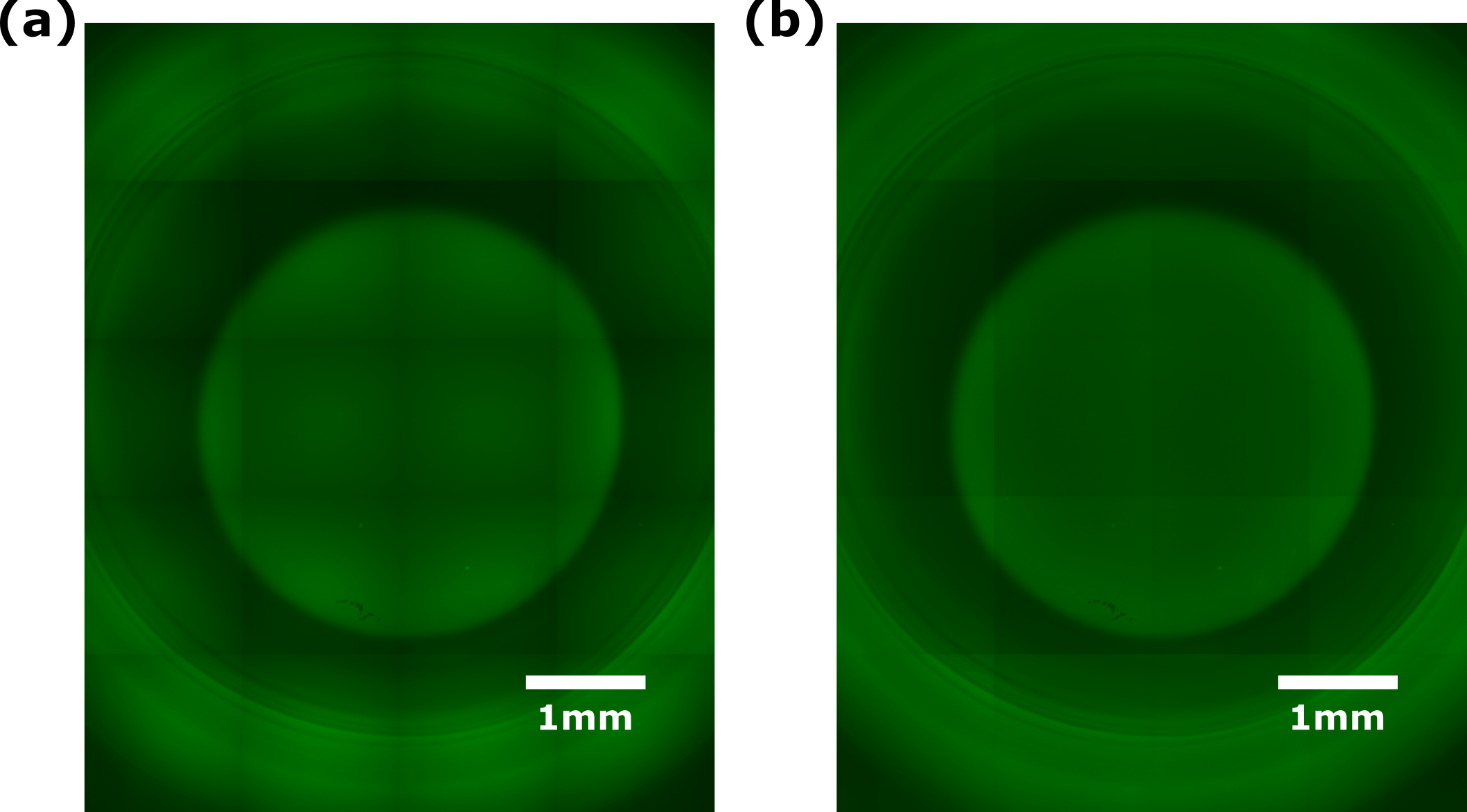}
    \caption{\textbf{Fluorescence images of ligand coverage of a well, before and after intensity normalisation.} The ligands constructs have been functionalised with the fluorophore Alexa Fluor 488, to enable visualisation of the DNA coverage. As described in the experimental methods section, a small circular region of the well has been functionalised with ligands, resulting in a steep ligand-density gradient at the perimeter of this region. The images are composed of multiple fields of view stitched together. Due to non-uniform illumination of the samples, the edges of the field of view have reduced fluorescence intensity, as can be seen in (a) with the images prior to normalisation. To account for this non-uniform illumination, the image intensities were normalised by scaling each field of view by a control image of a calcein solution; (b) shows the images post-normalisation.}
    \label{fig:image_normalisation}
\end{figure}

\clearpage

\begin{figure}
    \centering
    \includegraphics{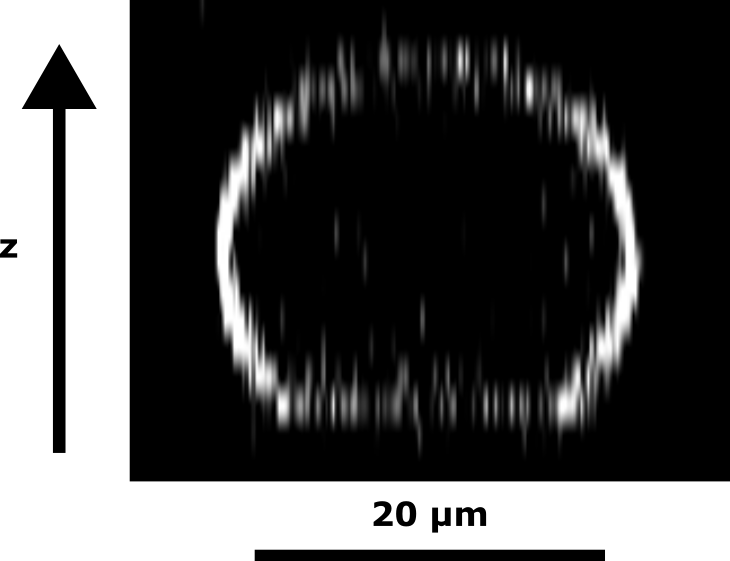}
    \caption{\textbf{Cross-sectional image of a vesicle adhered to a substrate.} The vesicle membrane has been functionalised with receptors, which bind to the complementary ligands anchored to a substrate. The lipid membrane of the vesicle also features a small proportion of fluorescently-labelled lipid molecules, allowing for the vesicle to be imaged with confocal microscopy. This cross-sectional view has been constructed from a confocal z-stack of the vesicle, and shows how the vesicle shape has been deformed to a truncated sphere with a flat contact region adhered to the substrate. Note that the diameter of the contact region is comparable (but slightly smaller than) the diameter of the vesicle; in our simulations and theoretical derivations we approximate the vesicle shape as a hemisphere.}
    \label{fig:vesicle_truncsphere}
\end{figure}

\clearpage

\clearpage

\begin{figure}
    \centering
    \includegraphics{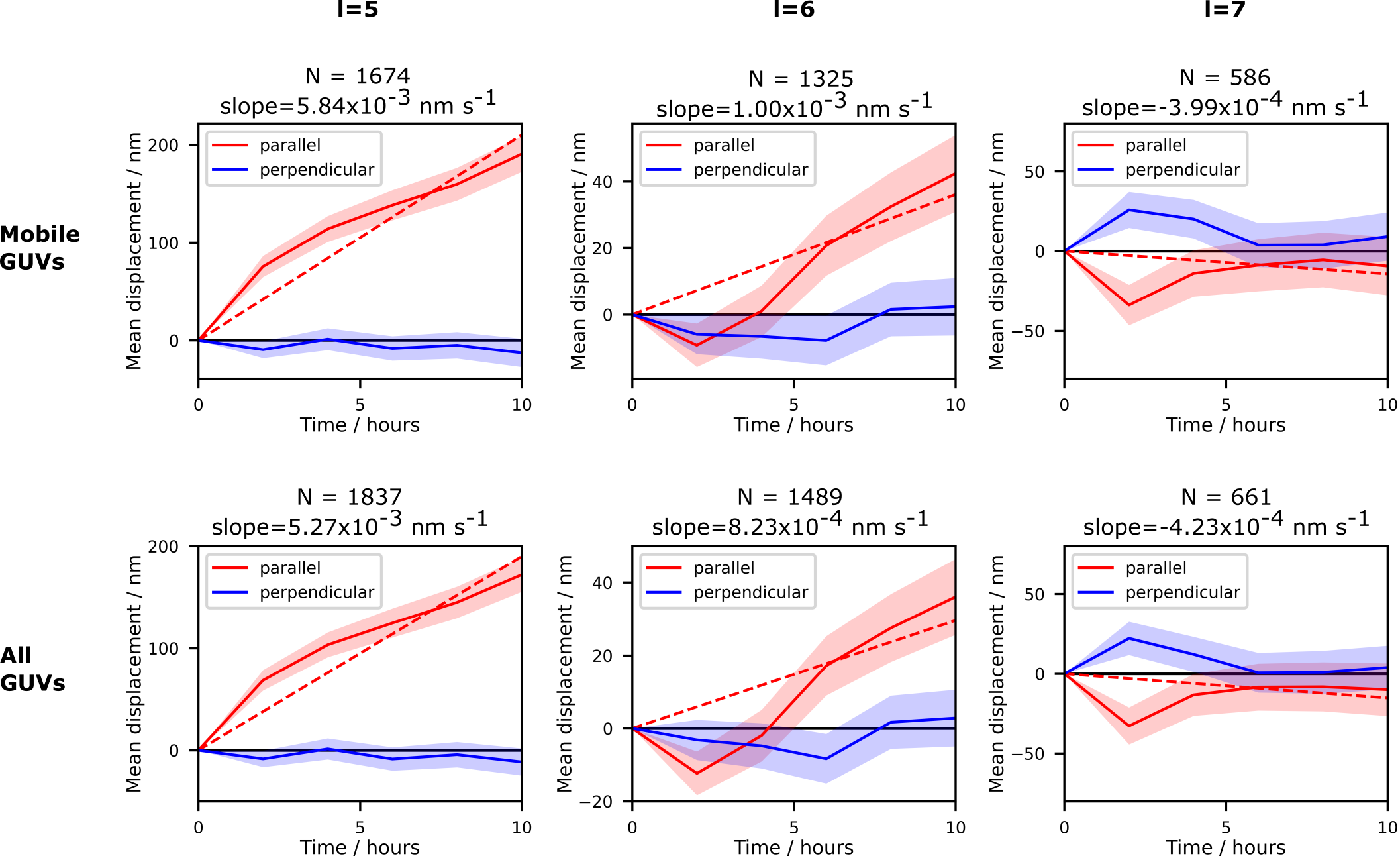}
    \caption{\textbf{Comparing experimental vesicle trajectories over time, for sticky end lengths $l=$~5, 6 and 7\,nt.} The line plots illustrate the mean displacements of vesicles parallel and perpendicular to the gradient direction, where positive values of displacement parallel to the gradient indicate vesicle motion towards higher ligand density regions. Straight lines have been fitted to the data for motion parallel to the gradient, to extract the average vesicle drifting velocities. Each plot is annotated with the number $N$ of vesicles averaged over, and the slope of the linear fit. Data are displayed for the filtered (mobile) vesicle population (top, also shown in Fig.~2) and for all vesicles (bottom); the filtering method is outlined in Section~\ref{subsec:msd} of the SI.}
    \label{fig:stickylength_dispvstime_exp}
\end{figure}

\clearpage

\begin{figure}
    \includegraphics{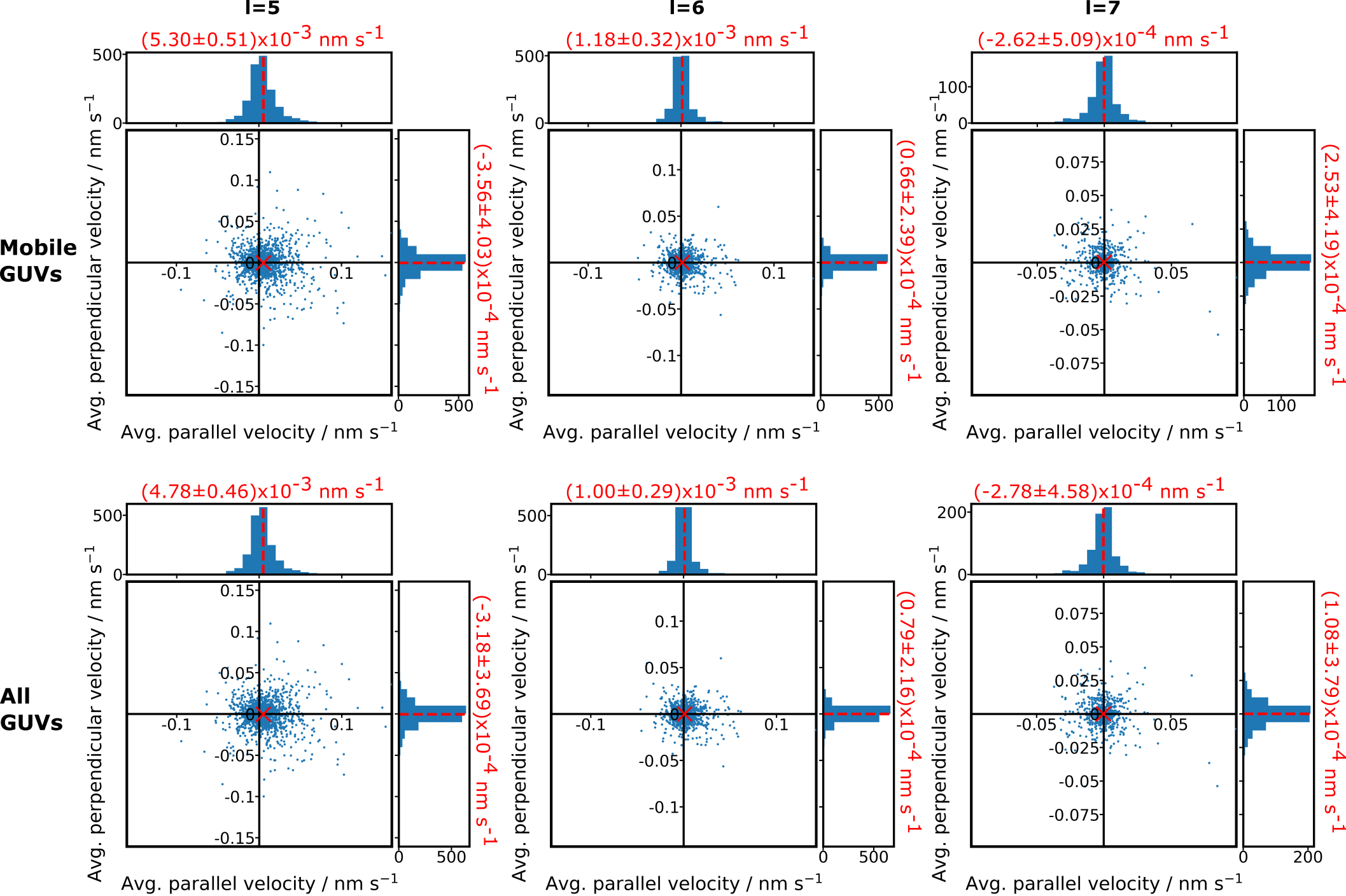}
    \caption{\textbf{Full distributions of velocities relative to the gradient direction, for vesicles measured in experiments.} The plots show experimental data for the distribution of vesicle velocities parallel and perpendicular to the gradient direction, where the velocity has been calculated as total displacement over the trajectory divided by the total duration. The data are for $l=$~5, 6 and 7\,nt sticky end systems, and includes all outliers. Positive values of displacement parallel to the gradient indicate vesicle motion in the direction of increasing ligand density. Data are displayed for the filtered (mobile) vesicle population (top, also shown in Fig.~2) and all vesicles (bottom); the filtering method is outlined in Section~\ref{subsec:msd}}
    \label{fig:stickylength_scatterandhist_exp}
\end{figure}

\clearpage

\begin{figure}
    \includegraphics{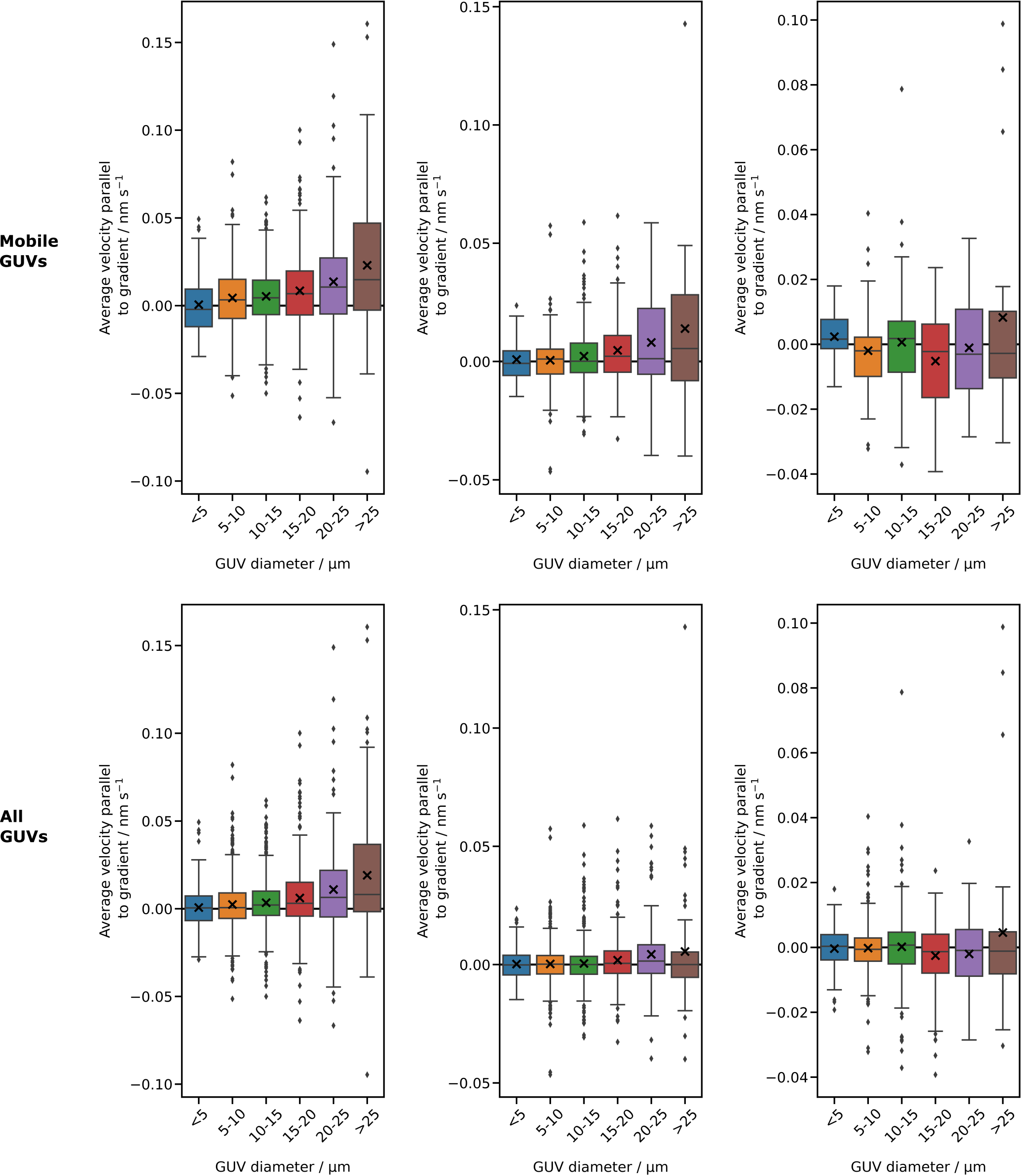}
    \caption{\textbf{Distribution of average vesicle velocity parallel to the gradient binned by vesicle size, for vesicles measured in experiments.} Data are shown for $l=$~5, 6 and 7\,nt sticky end systems, and the distributions are shown for the filtered (mobile) vesicle population (top, also shown in Fig.~3) and for all vesicles (bottom); the filtering method is detailed in Section~\ref{subsec:msd} of the SI. The average vesicle velocities have been calculated as total displacement divided by the trajectory duration, where positive velocity values indicate motion along the gradient in the direction of increasing ligand density. The mean of the distributions for each vesicle size range is marked with a cross, and all outliers have been included.}
    \label{fig:GUVsize_exp_par}
\end{figure}

\clearpage

\begin{figure}
    \includegraphics{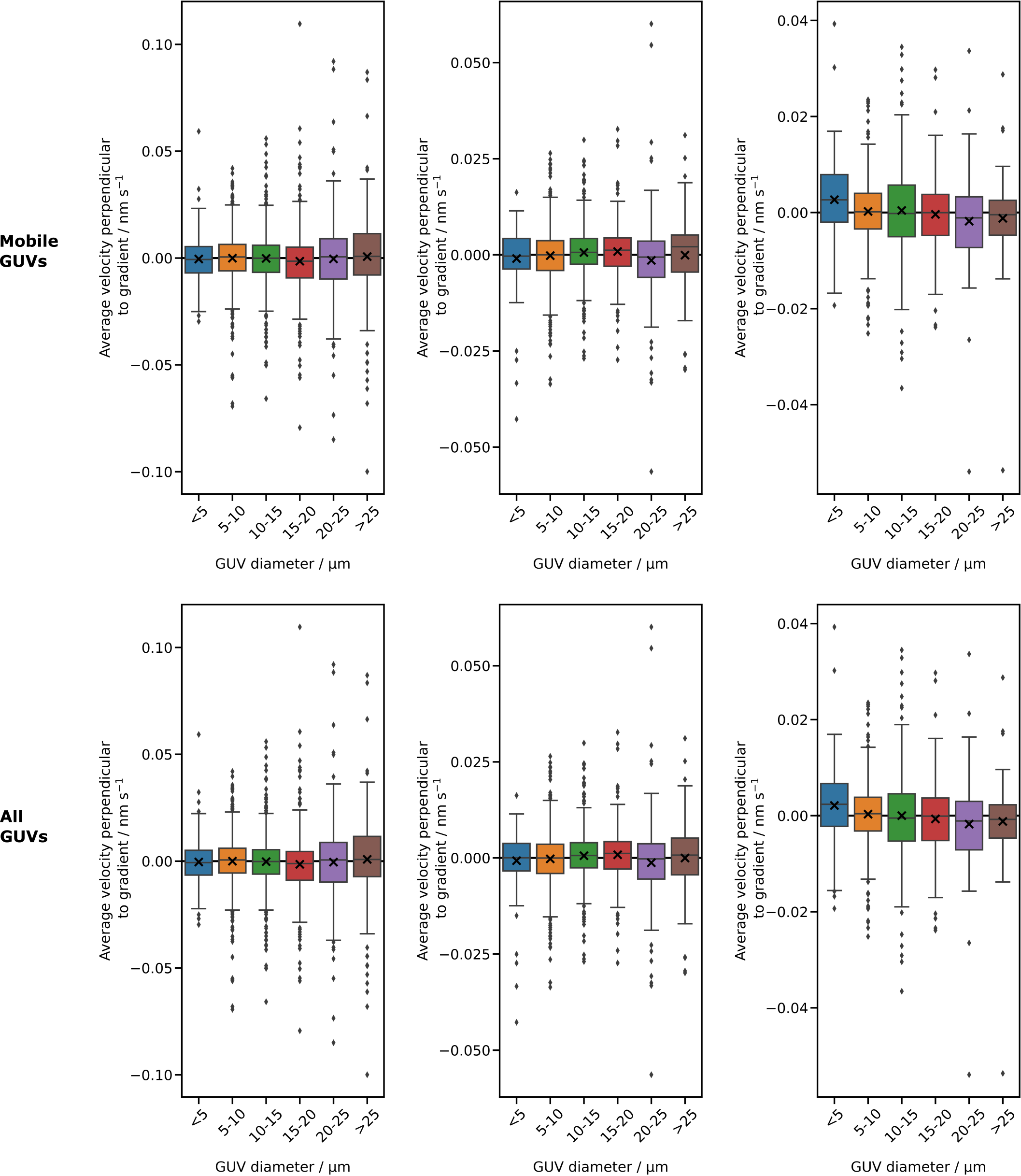}
    \caption{\textbf{Distribution of average vesicle velocity perpendicular to the gradient binned by vesicle size, for vesicles measured in experiments.} Data are shown for $l=$~5, 6 and 7\,nt sticky end systems, and the distributions are shown for the filtered (mobile) vesicle population (top) and for all vesicles (bottom); the filtering method is detailed in Section~\ref{subsec:msd} of the SI. The average vesicle velocities have been calculated as total displacement perpendicular to the gradient divided by the trajectory duration.The mean of the distributions for each vesicle size range is marked with a cross.}
    \label{fig:GUVsize_exp_perp}
\end{figure}

\clearpage

\begin{figure}[H]
    \centering
    \includegraphics{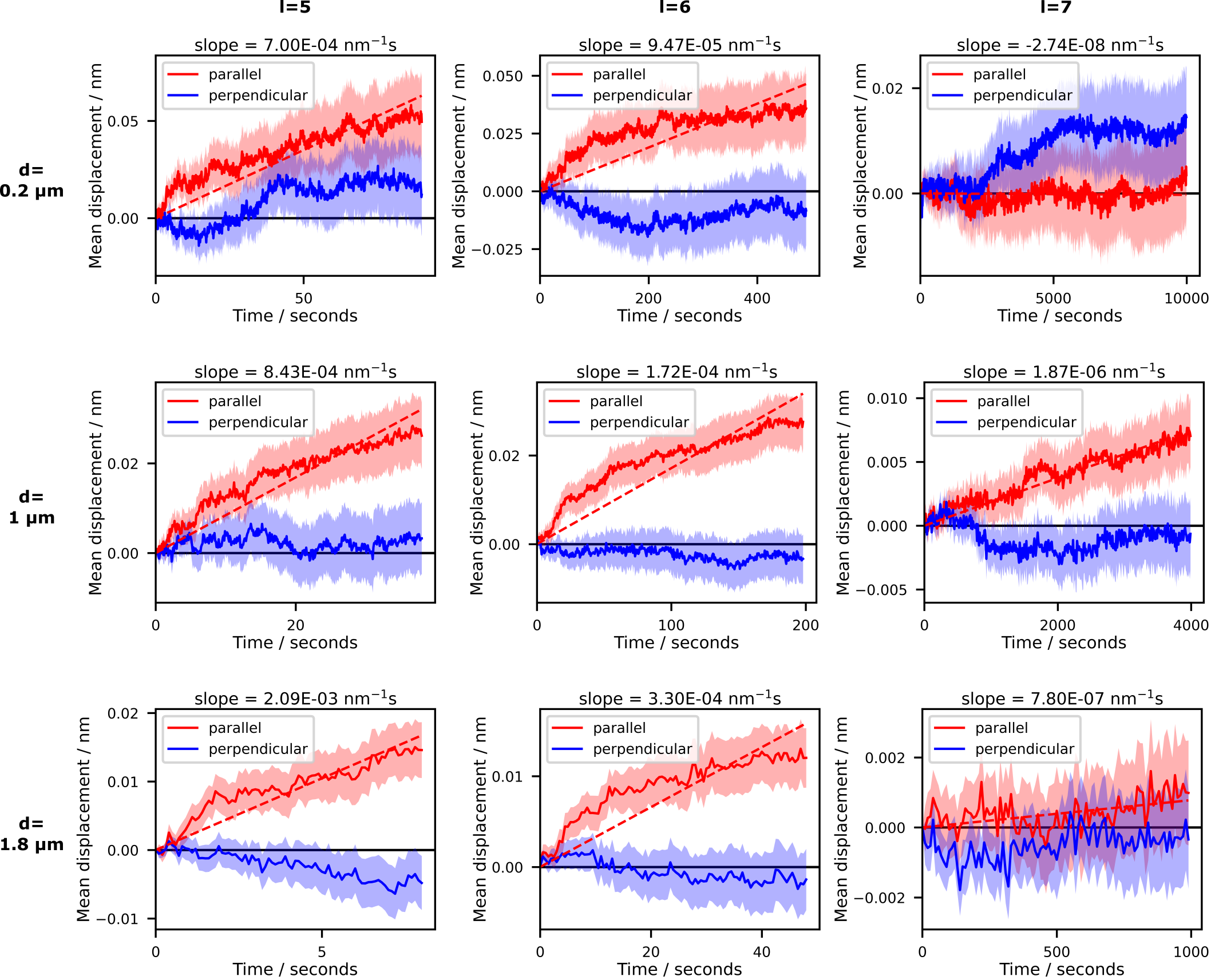}
    \caption{\textbf{Full trajectory data from simulations, for varying vesicle diameter and sticky end length.} The plots illustrate simulation results for mean vesicle displacement versus time, where the displacement has been decomposed into its components parallel and perpendicular to the gradient direction. The standard error of the mean displacement has been shaded. The data is for $l=$~5, 6, and 7\,nt sticky end systems, with vesicle diameters $d=$~0.2, 1 and 1.8 $\mu$m. Positive values of displacement parallel to the gradient indicate vesicle motion in the direction of increasing DNA surface density on the substrate. Straight lines have been fitted to the data for motion parallel to the gradient, and each plot has been annotated with the slope of the linear fit. Data for $d=1\,\mu$m are also shown in Fig.~2.}
    \label{fig:simulation_data_uncropped}
\end{figure}

\clearpage

\begin{figure}[!h]
    \includegraphics{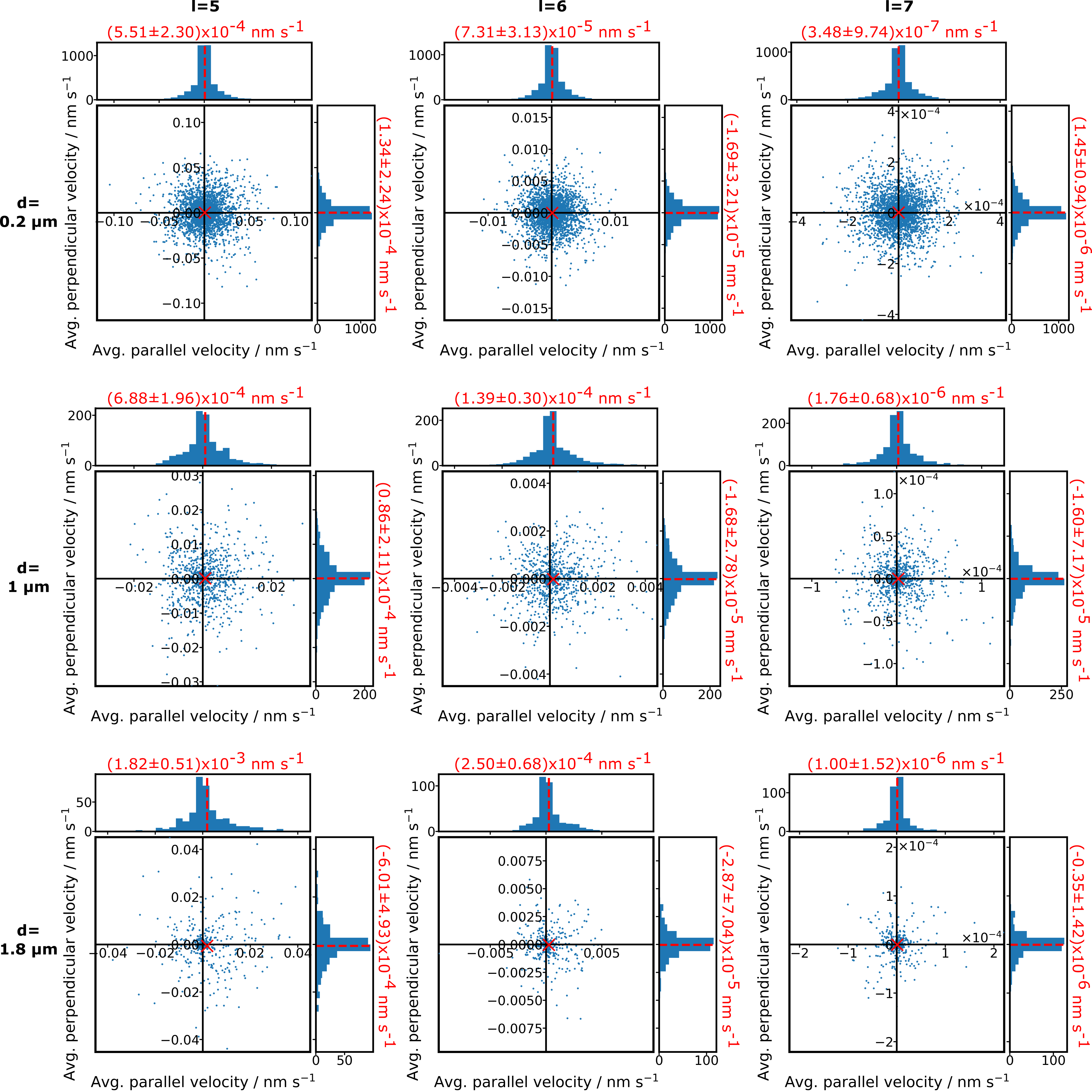}
    \caption{\textbf{Full distributions of simulated vesicle velocities relative to the gradient, for varying vesicle diameter and sticky end length.} The plots illustrate simulation results for vesicle velocity parallel and perpendicular to the gradient direction, where the velocity has been calculated as total displacement over the trajectory divided by the total duration. The data is for $l=$~5, 6, and 7\,nt sticky end systems, with vesicle diameters $d=$~0.2, 1 and 1.8\,$\mu$m. Positive values of displacement parallel to the gradient indicate vesicle motion in the direction of increasing DNA surface density on the substrate. The averages of the distributions are marked by red dashed lines and crosses. Data for $d=1\,\mu$m are also shown in Fig.~2.}
    \label{fig:simulation_data_scatandhist_uncropped}
\end{figure}

\clearpage

\begin{figure}[H]
    \centering
    \includegraphics{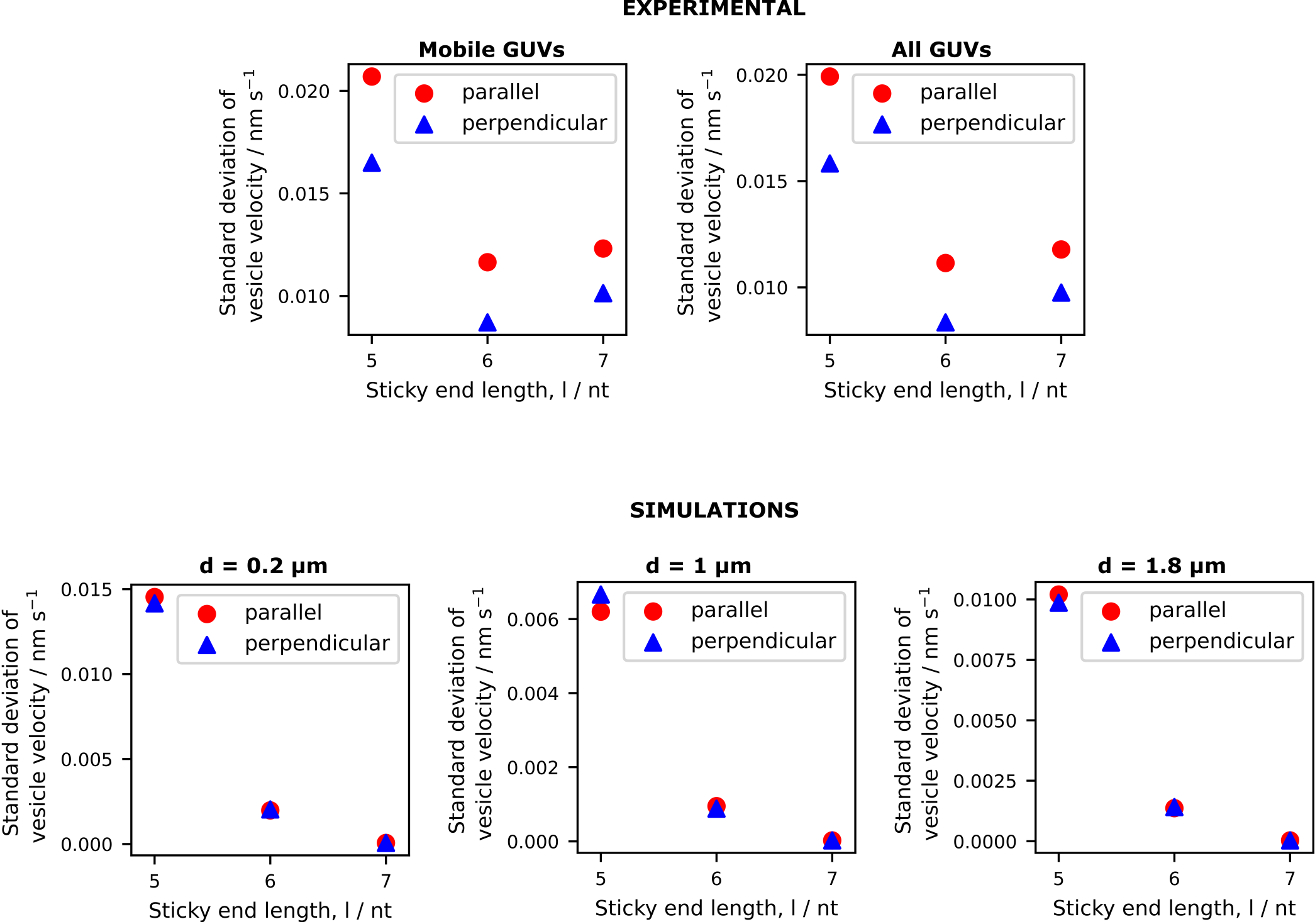}
    \caption{\textbf{Standard deviation of vesicle velocities versus sticky end length.} Experimental data (top) are shown for the filtered (mobile) vesicle population (top left) and for all vesicles (top right); the filtering method is outlined in Section~\ref{subsec:msd} of the SI. Simulation data are shown for vesicles with diameter $d=$~0.2\,$\mu$m, 1\,$\mu$m and 1.8\,$\mu$m.}
    \label{fig:velstd_vs_l}
\end{figure}

\clearpage

\clearpage

\begin{figure}[H]
    \centering
    \includegraphics{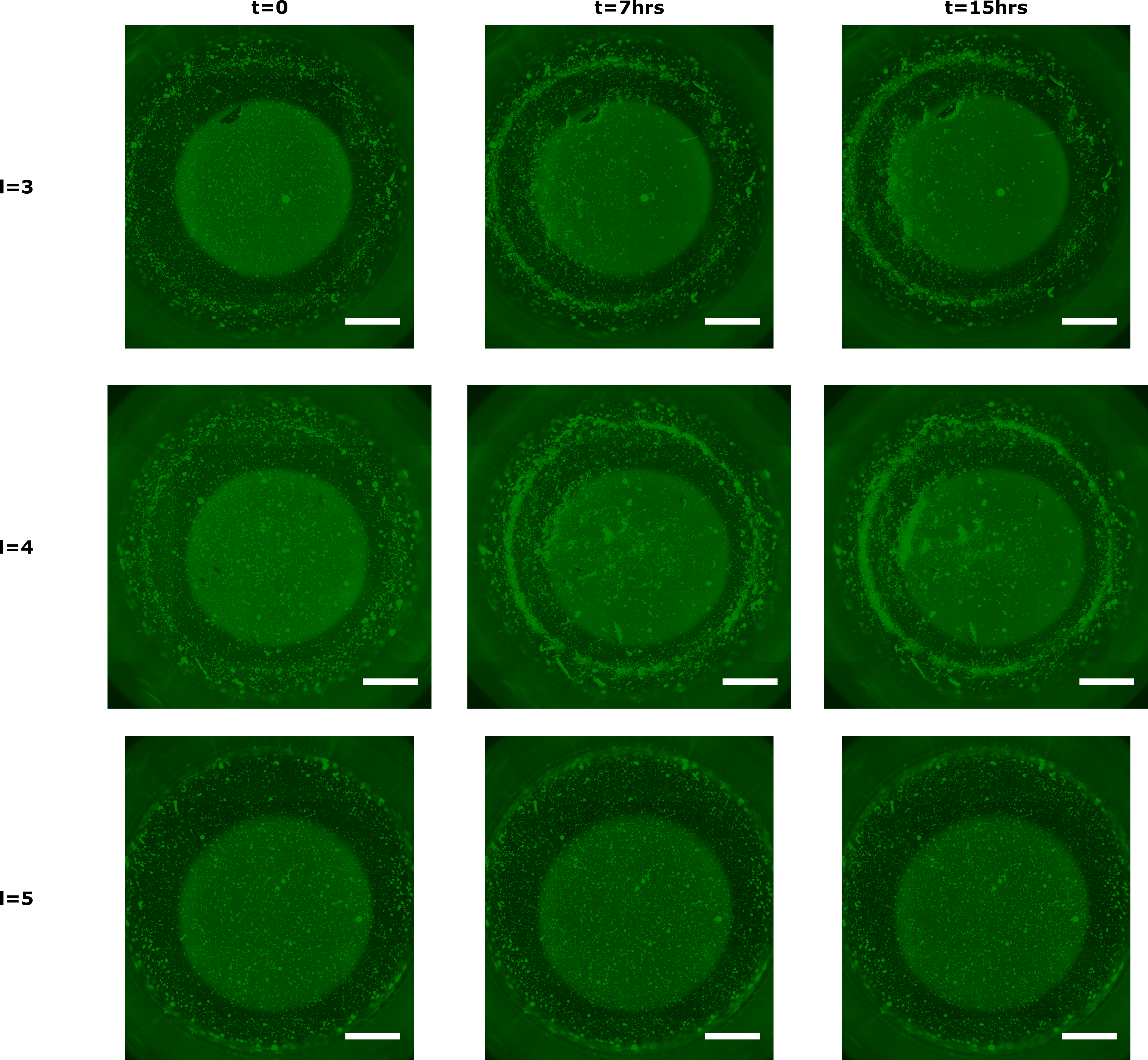}
    \caption{\textbf{Effects independent of surface adhesion cause vesicles to globally drift towards one side of the well in weakly adhering systems.} Microscopy images are shown for systems with sticky end lengths $l=$~3--5\,nt, at time $t=$~0, 7 and 14 hours after the experiment started. For the weakly binding systems $l=$~3 and 4, the majority of vesicles drift in one direction and move distances comparable to the width of the well. Scale bars correspond to a length of 1\,mm.}
    \label{fig:convective_flow}
\end{figure}


\clearpage

\begin{figure}
    \centering
    \subfloat{\includegraphics[width=65mm]{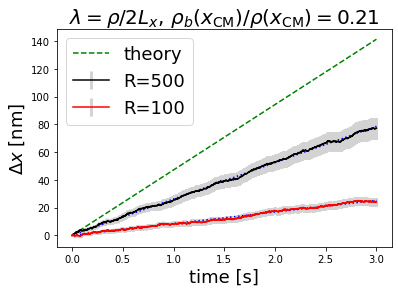}}
    \subfloat{\includegraphics[width=65mm]{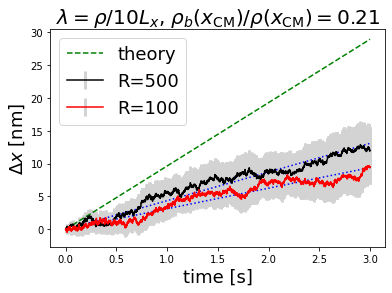}}
    \hspace{0mm}
    \subfloat{\includegraphics[width=65mm]{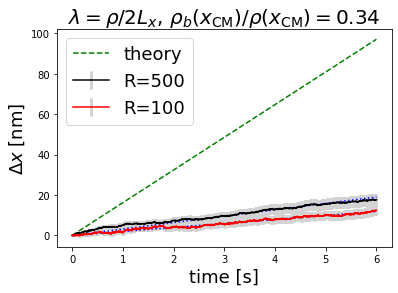}}
    \subfloat{\includegraphics[width=65mm]{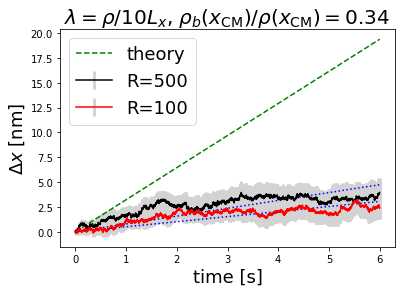}}
    \hspace{0mm}
    \subfloat{\includegraphics[width=65mm]{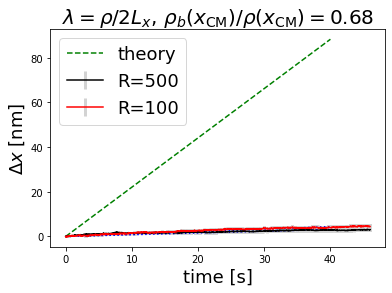}}
    \subfloat{\includegraphics[width=65mm]{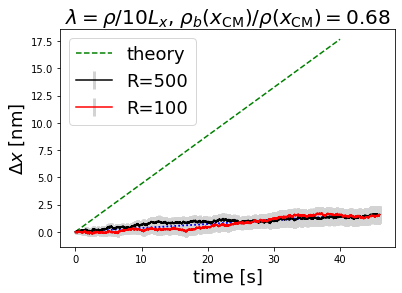}}
    
    \caption{\textbf{Comparison of simulated and theoretical vesicle displacements over time, parallel to the gradient, for varying gradients and bridge densities.} Mean displacements for varying gradients ($\lambda$) and density of bridges ($\rho_b$) predicted by simulations and by theory (Eq.~\ref{Eq:vdrift_lin}) are shown. In each plot, simulation results for two vesicle sizes (radius $R=$100\,nm and 500\,nm) are displayed, and straight lines have been fitted to them. $x$ refers to the distance along the gradient, and $\rho(x_{CM})$ is the ligand density within the contact region between the vesicle and substrate. $\rho$ and $L_x$ are the reference density and gradient length scale. The bridge density has been tuned by varying the off rates of the sticky ends. We find that when increasing $\rho_b$ the gap between simulation and theoretical predictions drastically increases.} 
    
    \label{Fig:nb}
\end{figure}

\clearpage

\begin{figure}
    \centering
    \subfloat{\includegraphics[width=65mm]{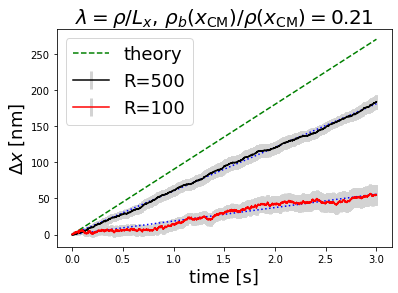}}
    \subfloat{\includegraphics[width=65mm]{SI_figs/FigureS18_1.png}}
    \hspace{0mm}
    \subfloat{\includegraphics[width=65mm]{SI_figs/FigureS18_2.png}}
    \subfloat{\includegraphics[width=65mm]{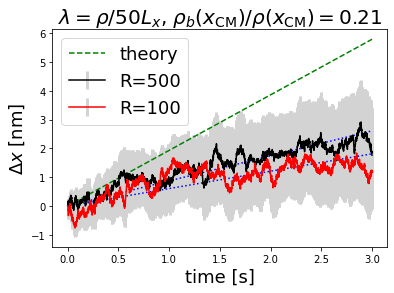}}
    
    \caption{\textbf{Comparison of simulated and theoretical vesicle displacements over time, parallel to the gradient, for varying gradients.} Mean average displacements for different gradients ($\lambda$) as predicted by simulations of vesicles with two different radii ($R$), and by theory (Eq.~\ref{Eq:vdrift_lin} in the main text). For small values of $\lambda$ the gap between theory and simulations decreases, and the drifting velocities no longer appear to depend on vesicle size (in agreement with Eq.~\ref{Eq:vdrift_lin}).}
    
    \label{Fig:lambda}
\end{figure}

\clearpage

\begin{figure}[H]
    \includegraphics{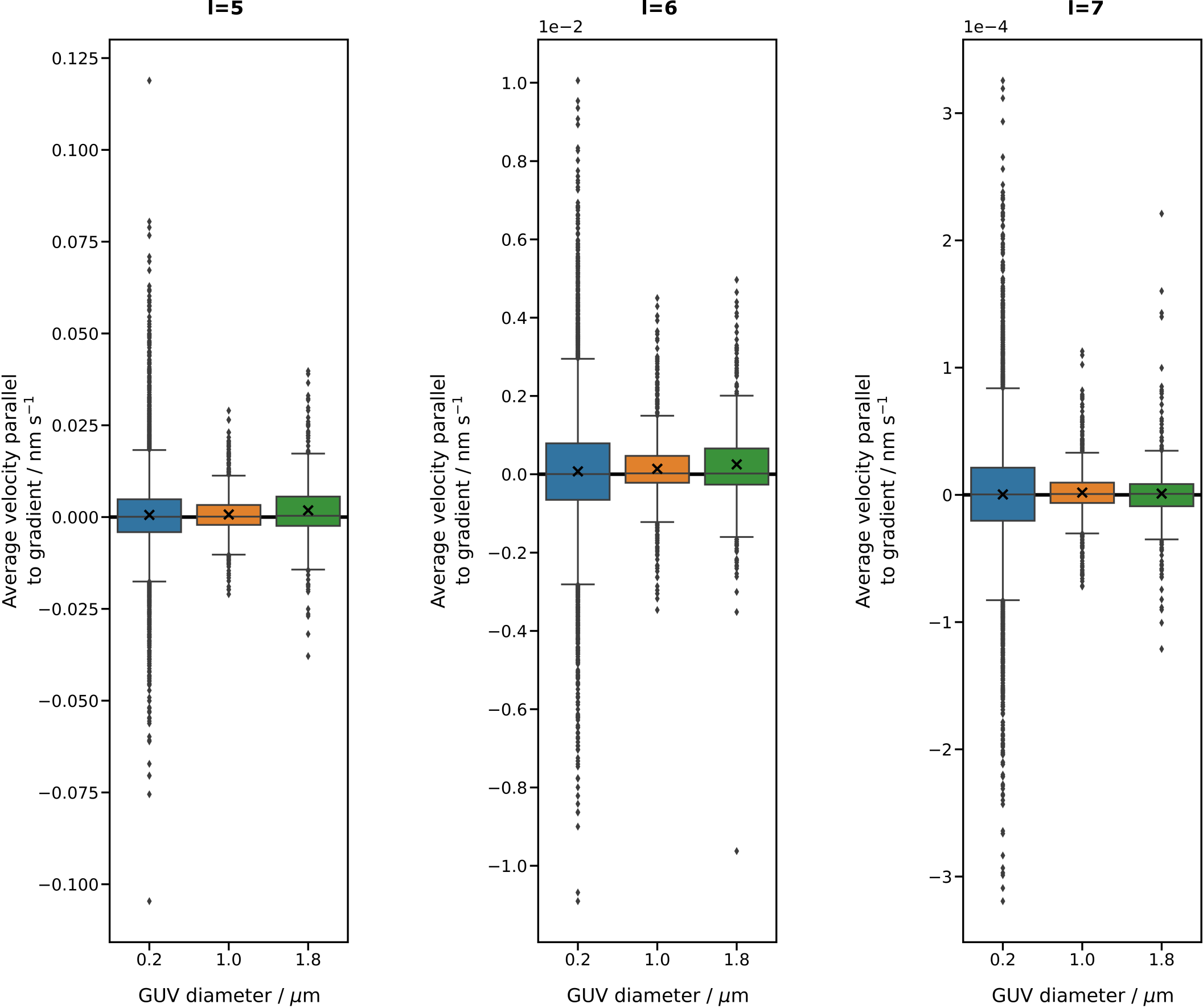}
    \caption{\textbf{Distributions of average vesicle velocity parallel to the gradient for 0.2\,$\mu$m, 1\,$\mu$m and 1.8\,$\mu$m diameter vesicles, from simulations.} Data for $l=$~5--7\,nt sticky end systems are plotted, including all outliers. Data are shown excluding outliers in Fig.~3. The average vesicle velocities have been calculated as total displacement parallel to the gradient divided by the trajectory duration; positive values of velocity indicate motion along the gradient in the direction of increasing ligand density on the substrate. The mean of the distributions for each vesicle size range is marked with a cross.}
    \label{fig:GUVsize_sim_par}
\end{figure}

\clearpage

\begin{figure}[H]
    \includegraphics{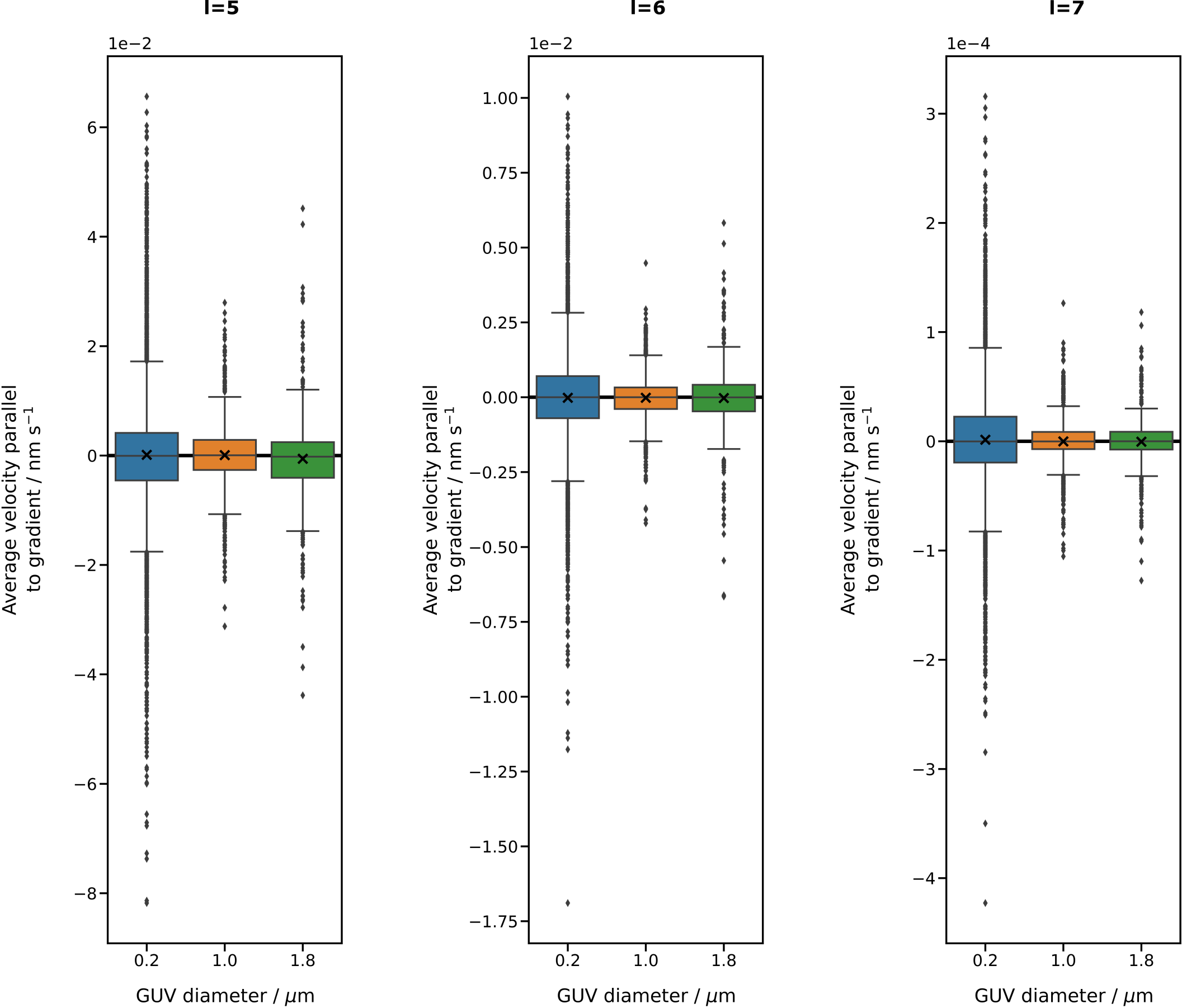}
    \caption{\textbf{Distributions of average vesicle velocity perpendicular to the gradient for 0.2\,$\mu$m, 1\,$\mu$m and 1.8\,$\mu$m diameter vesicles, from simulations.} Data for $l=$~5--7\,nt sticky end systems are plotted, including all outliers. The average vesicle velocities have been calculated as total displacement perpendicular to the gradient divided by the trajectory duration. The mean of the distributions for each vesicle size range is marked with a cross.}
    \label{fig:GUVsize_sim_perp}
\end{figure}

\newpage
\bibliography{SI_biblio}